\documentclass[letterpaper,12pt]{extarticle}  

\usepackage{import}
\usepackage{local}
\usepackage[left=.75in,top=.75in,bottom=.75in,right=.75in]{geometry}

\usepackage[utf8]{inputenc}
\usepackage[T1]{fontenc}

\usepackage{float}
\usepackage{physics}
\usepackage{graphicx}
\usepackage{dcolumn}
\usepackage{bm}
\usepackage{mathrsfs}
\usepackage{amssymb}
\usepackage{amsmath}
\usepackage[normalem]{ulem}
\usepackage{wrapfig}
\usepackage[compress,square,sort,comma,numbers]{natbib}

\usepackage{hyperref}
\usepackage{bookmark}

\usepackage{booktabs}
\usepackage[normalem]{ulem}
\useunder{\uline}{\ul}{}
\usepackage{lscape}
\usepackage{array}
\newcolumntype{L}[1]{>{\raggedright\arraybackslash}p{#1}}
\usepackage{rotating}
\usepackage{multicol}
\usepackage{caption}
\input{insbox}

\hypersetup{%
  breaklinks=true,
  colorlinks=true,
  linkcolor=black,
  filecolor=black,
  urlcolor=black,
  citecolor=black,
  pdfborder=0 0 0,
}

\begin{document}

\title{Learning functional groups in complex microbiomes}

\author{
\textbf{
\\
Matthew S. Schmitt\textcolor{Accent}{\textsuperscript{1,2*}}, %
Kiseok K. Lee\textcolor{Accent}{\textsuperscript{3,4,5*}}, %
Freddy Bunbury\textcolor{Accent}{\textsuperscript{3,4,5}}, %
Joseph A. Landsittel\textcolor{Accent}{\textsuperscript{6}}, %
Vincenzo Vitelli\textcolor{Accent}{\textsuperscript{1,2,5,7,8}**}, %
Seppe Kuehn\textcolor{Accent}{\textsuperscript{3,4,5,8**}} }\\
\begin{small}
\textcolor{Accent}{\textsuperscript{1}}James Franck Institute, University of Chicago, Chicago, Illinois 60637, U.S.A. \\ 
\textcolor{Accent}{\textsuperscript{2}}Department of Physics, University of Chicago, Chicago, IL 60637, U.S.A. \\
\textcolor{Accent}{\textsuperscript{3}}Department of Ecology and Evolution, University of Chicago, Chicago, IL 60636, U.S.A. \\
\textcolor{Accent}{\textsuperscript{4}}Center for the Physics of Evolving Systems, University of Chicago, Chicago, IL 60636, U.S.A. \\
\textcolor{Accent}{\textsuperscript{5}}Center for Living Systems, University of Chicago, Chicago, IL 60636, U.S.A. \\
\textcolor{Accent}{\textsuperscript{6}}Department of Engineering Sciences and Applied Mathematics, Northwestern University, Evanston, IL, U.S.A \\
\textcolor{Accent}{\textsuperscript{7}}Leinweber Institute for Theoretical Physics, University of Chicago, Chicago, IL 60637, U.S.A. \\
\textcolor{Accent}{\textsuperscript{8}}National Institute for Theory and Mathematics in Biology, Northwestern University and University of Chicago, Chicago, IL, U.S.A. \\
\textcolor{Accent}{\textsuperscript{*}} These authors contributed equally \\ 
\textcolor{Accent}{\textsuperscript{**}}Correspondence: \textcolor{Accent}{vitelli@uchicago.edu,
seppe.kuehn@gmail.com} \\ \end{small}
}

\maketitle

\begin{linenumbers}

\section{abstract}

From soil to the gut, communities composed of thousands of microbes perform functions such as carbon sequestration and immune system regulation. Here, we introduce a data-driven approach that explains how community function can be traced to just a few groups of microbes or genes.
In gut communities, our neural-network based clustering algorithm correctly recovers known functional groups. 
In the ocean metagenome, it distills $\sim$500 gene modules down to three sparse groups highlighting survival strategies at different depths. 
In soils, it distills $\sim$4400 bacterial species into two groups that enter a mathematical model of nitrate metabolism. By combining interpretable ML with strain isolation and sequencing experiments, we connect the metabolic specialization of each group to community-wide responses to perturbations. This integrated approach yields simple structure-function maps of microbiomes, allowing the discovery of molecular mechanisms underlying human and environmental health. More broadly, we illustrate how to do function-informed dimensionality reduction in biology.

\newpage

\clearpage
\newpage

\begin{figure}{tp}
    \centering
    \includegraphics[width=0.6\textwidth]{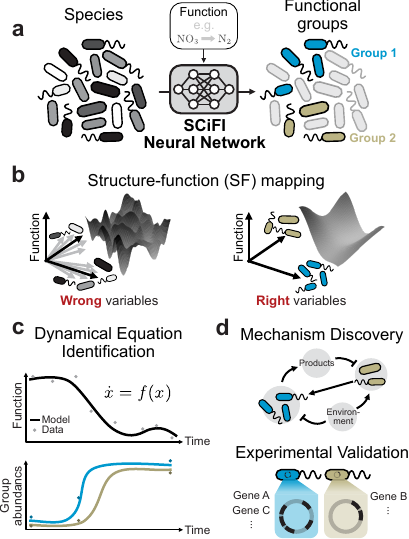}
    \caption*{
    \textbf{Graphical Abstract: An integrated ML and experimental pipeline to discover functional groups and their dynamics in complex microbiomes and beyond.}
    (a) First, our Soft Clustering Function Informed (SCiFI)  algorithm identifies functional groups directly from species abundances data using neural networks. Crucially, the learned functional groups are informed by a chosen community function.
    (b) Descriptions of the system  in terms of the original high-dimensional abundances of individual species are prohibitively complex (left). By contrast, a description in terms of the few functional groups identified by SCiFI leads to a simple structure-function map (right). 
    (c) Next, the identified groups can be directly input as variables into predictive mathematical models for the dynamics of the community.
    (d) The last step of our pipeline relies on the identified groups comprising only a small number of species. This sparsity enables targeted experiments that interrogate isolated species (e.g. with whole-genome sequencing or phenotyping) shedding light on the mechanistic underpinnings of the structure-function map with potential applications beyond microbiomes, from gene expression to neuronal activity.
    }
    \label{fig1:intro}
\end{figure}

\clearpage
\newpage

\begin{figure}[th]
    \centering
    \includegraphics[width=1.0\textwidth]{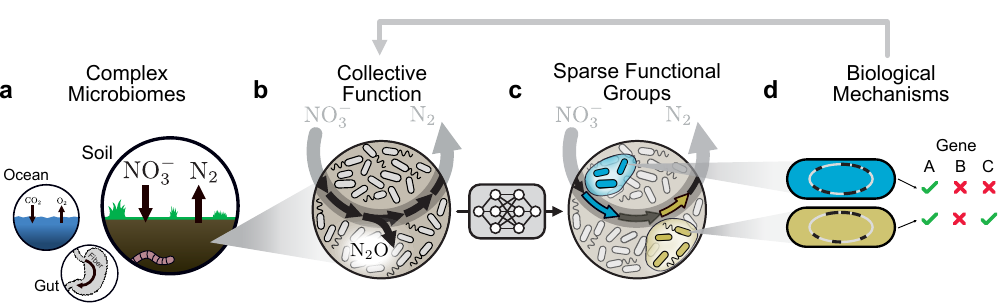}
    \caption{
    \textbf{Data-driven discovery of functional groups and their dynamics}
    (a) Microbial communities perform crucial environmental functions from the soil to the ocean to the gut.
    (b) In soils, microbes collectively reduce nitrate to dinitrogen gas in a process called denitrification. This process is composed of several discrete steps (black arrows), one of which produces the potent greenhouse gas nitrous oxide.
    (c) Our machine learning method, SCiFI, automatically finds the  functional groups of bacteria (blue, green) that contribute to community function. These groups correspond to distinct metabolic functions (arrows).
    (d) By sequencing the genomes of group members, we can disentangle the biological mechanisms leading to collective function.
    }
    \label{fig1}
\end{figure}

\section{Introduction}

Bacteria account for $\sim$10\% of all biomass on earth \cite{bar-on_phillips_biomass}. 
Collectively, they perform important metabolic functions such as transferring gigatons of carbon \cite{warner_spatial_2019_carbon,bondlamberty_globally_2018_carbon,hashimoto_divergent_2023} and nitrogen \cite{kuypers_microbial_2018} between the soil and the atmosphere every year, clearing pollutants from wastewater~\cite{third2001canon}, and protecting against pathogen colonization \cite{spragge_microbiome_2023}. 
These functions arise from the concerted metabolic activity of structured communities consisting of hundreds or thousands of individual species.
This complexity makes the corresponding structure-function mapping difficult to decipher.

However, the biochemistry of metabolism suggests that the composition of a bacterial community and its relationship to function can be described in terms of just a few functional groups that perform distinct roles \cite{winogradsky1887schwefelbacterien,flint_rumen_1997,enke_modular_2019,madigan2014brock,Strous1999_guilds,rawlings_heavy_2002_guilds,bryant_methanobacillus_1967_guilds,daims_complete_2015_guilds}.
How can we distill these groups automatically from data and mechanistically relate them to community function?

As an example, consider the microbiome function triggered when you put fertilizer in damp soil (Fig.~\ref{fig1}a). 
The metabolic process of denitrification, responsible for significant greenhouse gas emissions, rapidly ensues \cite{firestone_temporal_1979,bremner_nitrous_1978}. 
The greenhouse gas nitrous oxide (N$_2$O) is emitted by the collective action of functional groups, each specialized in specific steps of the metabolic cascade, e.g. conversion of NO$_3$ (fertilizer) to N$_2$O or of N$_2$O to N$_2$ (Fig.~\ref{fig1}b) \cite{pold_denitrification_2025}. 
Identifying functional groups, and their role in the community, has traditionally required extensive experimental studies of isolates and their interactions \cite{gamble_numerically_1977,crocker2024environmentally}.

Here, we introduce a machine-learning algorithm called SCiFI that discovers functional groups from data, yielding interpretable structure-function maps for complex microbiomes (Fig.~\ref{fig1}c). A key finding is that each functional group contains only a few relevant members. By experimentally targeting this subset of bacterial species we can rapidly evaluate the biological mechanisms defining each group's role in community function (Fig.~\ref{fig1}d). In soil, this integrated machine-learning and experimental approach reveals how the members of two genetically distinct groups collectively determine denitrification dynamics across varying environments.

\section{Results}
\subsection{A machine learning approach to function-informed clustering: the SCiFI algorithm}
Our goal is to distill a small number of functional groups from abundance measurements of thousands of microbial species in a way that is informed by community function. 
The task of aggregating species into groups is an instantiation of a more general one called dimensionality reduction: obtaining a low-dimensional representation of a complex system that retains its salient features.
Unlike physical systems that are often representable by a single set of collective variables governing their behavior (e.g. $P$, $V$, $T$ for a gas), biological systems do not typically possess a unique decomposition into collective variables.
In proteins, for instance, the groups of residues governing thermal stability are distinct from those governing catalytic activity \cite{halabi_2009} while in cells, the genes that control differentiation into fibroblasts are not those controlling differentiation into neurons \cite{vierbuchen_direct_2010}.
Similarly, in microbiomes the groups relevant for denitrification \cite{gamble_numerically_1977} differ from those relevant for methanogenesis \cite{jones_methanogens_1987}.
Nevertheless, with few exceptions \cite{Shan2023,Zhao2024}, dimensionality reduction in microbiomes is typically attempted without explicitly linking groups and function~\cite{faust_microbial_2012,kurtz_sparse_2015,Friedman2012}. Species are first clustered into groups via statistical or genomic methods, and only later the resulting groups are correlated with function \cite{gevers_treatment-naive_2014,deutschmann_disentangling_2021,zhou_gut_2022}.

Here, we introduce a machine learning algorithm that simultaneously identifies groups and associates them with collective function. Our soft clustering function-informed (SCiFI) algorithm exploits a Gumbel softmax trick that turns cluster identity (typically a categorical variable) into a continuous (i.e., soft) variable (Methods). This continuous representation, beyond enabling the model to express uncertainty, is fully differentiable. Hence, optimization can be performed with gradient descent which allows information to flow, via gradients, from function to clustering. 
This crucial feature is what allows SCiFI to find groups that are informative of specific functional processes.

\begin{figure}[tp]
    \centering
    \includegraphics[width=0.67\linewidth]{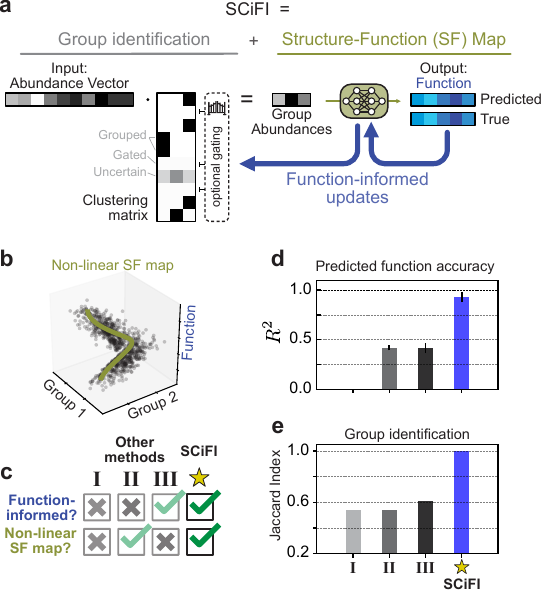}
    \caption{
    \textbf{
    SCiFI: a neural network-based method to identify functional groups}
    (a) Our pipeline consists of two steps: first, species abundances are aggregated via matrix multiplication with a grouping matrix; second, group abundances are used to predict the target function using a neural network. 
    An optional gating term allows entire rows of the grouping matrix to be set to zero (Methods).
    During training, the grouping matrix is learned simultaneously with the neural network using gradients from the loss function. 
    (b) A simple model of nonlinear data, in which the function depends on the inputs only through grouped abundances (Methods).
    (c) We compare our method to three alternative models which lack SCiFI's ability to capture a non-linear function map, find function-informed clusters, or both. Each method is described in the main text.
    (d) $R^2$ of function predictions on a held-out subset of the data for each method.
    (e) Group recovery for each method as measured by Jaccard Index (Methods).
    }
    \label{fig:the_model}
\end{figure}

Figure~\ref{fig:the_model}a describes SCiFI in a nutshell, emphasizing how it simultaneously yields the group identification (left) and the structure-function map (right). 
The input is community structure, represented as a species abundance vector (gray). The output is a prediction of function (blue), which may be a vector (e.g. a time series of metabolite concentration) or a scalar (e.g. the concentration at a single time point). 
To go from input to output, we first group species by multiplying the abundance vector with an interpretable clustering matrix: each row of this matrix encodes by construction the group assignment for one species (Methods). 
This aggregation of species by summation assumes that members within a group are indistinguishable.
In systems with a large number of species many are irrelevant for function prediction. In such cases, we encourage SCiFI to find sparse groups, i.e. comprising only a few species, with an optional gating step achieved via regularization and presented in details in Methods. 
The resulting group abundances are in turn fed to a neural network (green) that maps them to function.
The neural network enables SCiFI to accommodate nonlinear structure-function maps which, as we show below, occur in real systems. 
SCiFI updates both the clustering matrix and the neural network parameters (blue arrows) to minimize the deviation between predicted and true function. Simultaneous updating is the architectural feature that ensures feedback between clustering and structure-function mapping during training. 
This is how SCiFI's groups become function informed.

We benchmark SCiFI using a simulated dataset (Methods) where function is generated from a small number of groups through a nonlinear structure-function map (Fig.~\ref{fig:the_model}b). Nonlinearity is a typical feature of the experimental microbiome datasets we analyze below. For benchmarking we consider other models that can be used to find functional groups. Each of these models lack one or both of the SCiFI's defining features, namely its nonlinearity and function-informed clustering (Fig.~\ref{fig:the_model}c). 
Model I learns a grouping independent of function by first clustering with a co-occurrence network and then predicting the function from group abundances with linear regression \cite{gevers_treatment-naive_2014}.
Model II performs the same first abundance-based clustering step, but uses a neural network to learn the mapping between group abundances and function. 
Finally, Model III, introduced in Ref.~\cite{Zhao2024}, learns a function-informed clustering with a Monte Carlo approach combined with a linear structure-function mapping.

SCiFI accurately predicts function (Fig.~\ref{fig:the_model}d) and recovers functional groups as measured by the Jaccard score (Fig.~\ref{fig:the_model}e and SI~Figs.~\ref{si_fig:synthetic_data_benchmarking},~\ref{si_fig:linear_degradation_chain}), which computes the ratio of correct group classification to the total group size. 
Benchmarking reveals that these models generally fail to recover the correct groups and predict function (Fig.~\ref{fig:the_model}d-e).
Models I and II fail because they first misidentify groups, which then inhibits function prediction.
In Model III the situation is reversed: its inability to predict nonlinear function leads it to incorrectly identify groups (SI Fig.~\ref{si_fig:benchmark_linear_SF_map}).

\begin{figure}[tp]
    \centering
    \includegraphics[width=0.99\linewidth]{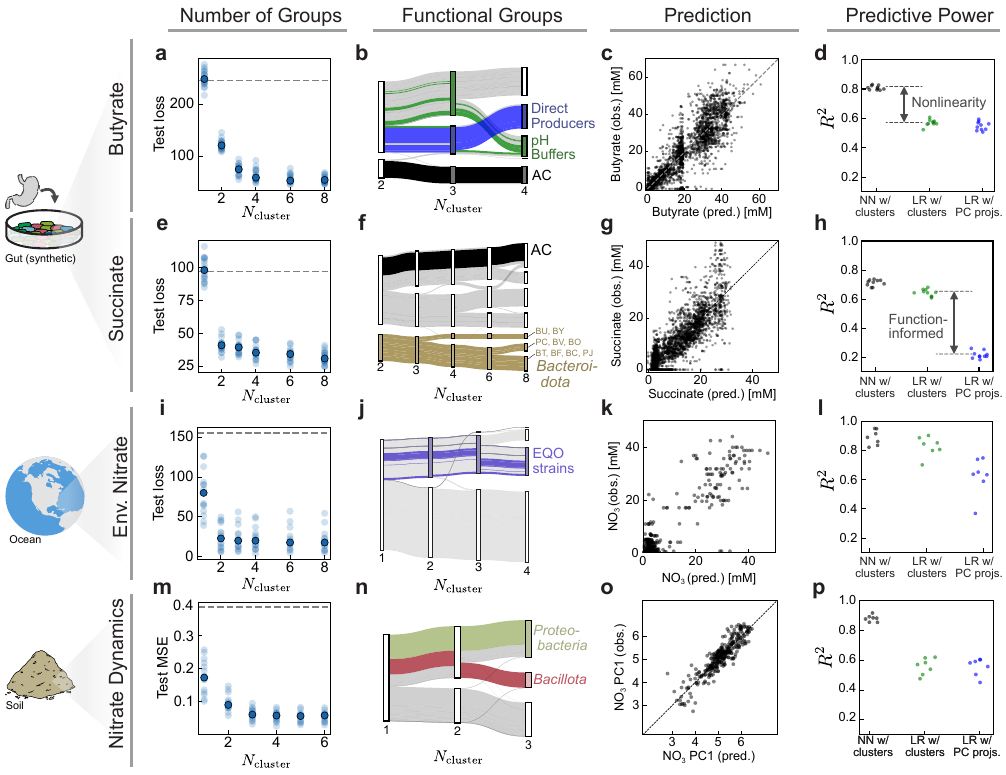}
    \caption{
    \textbf{SCiFI correctly learns functional groups in the gut, soil, and ocean microbiome}
    (a) We train our model to predict butyrate production based on the abundances of 30 bacterial strains in synthetic gut communities (data from Ref.~\cite{Clark2021}).
    Test loss (mean squared error) of predicted butyrate concentrations is shown for varying numbers of functional groups $N_{\text{cluster}}$. Each faint dot is one model trained with one particular test/train split of the data. The solid dot shows the median across the entire ensemble of models ($N=12$ test/train splits total). Dashed gray line shows the variance of butyrate as a reference.
    (b) Flow chart showing how group structure changes with varying $N_{\text{cluster}}$. Thickness of each bar corresponds to the average abundance of that species across the dataset. Several individual species are highlighted; for interpretation see the main text.
    (c) SCiFI predictions (with $N_{\text{cluster}}=4$) versus observed butyrate concentrations in each sample.
    (d) Comparison of test-set $R^2$ values for three models: the neural network which uses the identified cluster abundances as inputs; a linear regression which uses the cluster abundances as inputs; and a linear regression which uses the projection onto the first 4 principal components as inputs.
    Each point is a model trained and evaluated using a different test-train split of the data (Methods).
    All models use a 4-dimensional input (that is, either $N_{\text{clusters}}=4$ or $N_{\text{PCs}}=4$).
    (e-h) Same plots as in the top row, using succinate as the target function. The models in (g) and (h) use $N_{\text{clusters}}=2$.
    (i-l) Results for genus abundances in marine communities. Target function is a scalar measurement of environmental nitrate concentration.
    The models in (k) and (l) use $N_{\text{clusters}}=2$.
    (m-p) Results for phylum abundances in soil communities. Target function is a time-series of nitrate concentrations.
    The models in (o) and (p) use $N_{\text{clusters}}=3$.
    All of these models in this figure are trained without the optional gating step described in Figure~\ref{fig:the_model}a. For each dataset, a comparison to several other methods may be found in SI~Fig.~\ref{si_fig:fig3_method_comparison}.
    }
    \label{fig:validation}
\end{figure}

\subsection{SCiFI algorithm correctly identifies functional groups for \textit{in vitro} gut microbiomes}
We validate SCiFI's performance on synthetic gut microbiome experiments from Ref.~\cite{Clark2021}, which form communities consisting of combinations of 30 gut bacteria. This dataset consists of hundreds of paired measurements of strain abundances and fermentation products relevant for gut health including butyrate and succinate.

We first aim to identify functional groups relevant for butyrate production. 
The correct number of groups is \textit{a priori} unknown. To find it, we train multiple networks over a range of possible group numbers $N_{\text{groups}}$ and see when the error in butyrate prediction stops decreasing (Fig.~\ref{fig:validation}a).
Because the number of species is small, we do not require SCiFI's optional gating step.
We find that four functional groups are necessary to accurately predict butyrate production of the whole community. The groups we learn are robust (SI Figs.~\ref{si_fig:fig3_CV_consistency},~\ref{si_fig:fig3_hparam_consistency}) and interpretable in terms of the phenotypes described in Ref.~\cite{Clark2021} (Fig.~\ref{fig:validation}b and SI~Fig.~\ref{si_fig:fig3_specspec_corrs}).
The first group contains a single member, \textit{Anaerostipes caccae} (AC) a butyrate producer which can switch between low butyrate (sugar-utilizing butyrate production) and high butyrate (lactate-utilizing butyrate production) producing states depending on environmental pH~\cite{Clark2021}. This explains the second group of so-called ``pH buffers'', which are species that are highly correlated with pH that modulate butyrate production by impacting AC's mode of production (SI Fig.~\ref{si_fig:clark_correlations_and_avg_abd}a).
The third group contains the rest of the direct butyrate producers, while the final group contains the remaining species which are not important for butyrate production. 
These groups, which were independently experimentally confirmed in Ref.~\cite{Clark2021} show that SCiFI correctly learns functional groups in the microbiome.

Knowledge of these four coarse-grained groups suffices to accurately predict butyrate concentration (Fig.~\ref{fig:validation}c). This is possible only because SCiFI can learn non-linear structure-function relationships.
If the non-linearity is removed and we instead use linear regression to predict butyrate from the (learned) group abundances, the model suffers a significant decrease in accuracy (Fig.~\ref{fig:validation}d and SI~Fig.~\ref{si_fig:fig3_method_comparison}).

Repeating our analysis using succinate concentration instead of butyrate results in a completely different group structure (Fig.~\ref{fig:validation}e-f).
SCiFI correctly ignores the dominant species (measured by average abundance) AC, and instead isolates strains from the \textit{Bacteroidetes} phylum. In contrast to the butyrate case, the loss continues to see modest decreases as we increase the number of groups beyond the initial plateau (at $N_{\text{cluster}}=2$) which correspond to subsequent subdivisions of \textit{Bacteroidetes}.

\clearpage

How can we understand how the learned groups contribute to community function? 
To move forward we use the learned groups as the starting point for a genomic analysis.
Namely, we use sequenced genomes of group members to search for key enzymes that produce succinate using the Biocyc database~\cite{Qian2025}. 
This reveals that all nine strains assigned to the putative succinate-producing group (brown in Fig.~\ref{fig:validation}f) have at least one copy of the fumarate reductase gene.
By contrast, all other strains lack fumarate reductase except for the two species EL and FP, which have low average abundance (SI Fig.~\ref{si_fig:clark_correlations_and_avg_abd}b).

In contrast to the butyrate case, the relationship between the functional groups and succinate prediction is simple: a linear model predicts succinate from the group abundances nearly as well as the neural network (Fig.~\ref{fig:validation}g; green in Fig.~\ref{fig:validation}h).
However, this is only true when the clustering is function-informed.
If we instead project the abundance data onto its principal components (blue) as opposed to our function-informed clustering matrix, a linear model is incapable of predicting succinate concentration. 
This indicates that the dominant variance in abundances does not reflect changes in succinate concentration and highlights the importance of identifying clusters in a function-informed manner.

In summary, because it is function-informed, SCiFI can identify multiple different groupings from the same set of abundance data. 
Further, the groups that we learn have clearly interpretable biological roles, and in the case of succinate they also possess a distinct genetic signature. 

\subsection{SCiFI finds functional groups in natural ocean and soil microbiomes.}
In contrast to the synthetic communities above, in natural ocean and soil samples the experimentalist has little to no control over the composition of the community. 
Our method identifies relevant functional groups even in these settings.

First, we aim to find functional groups in the ocean microbiome using the Tara Oceans dataset \cite{Sunagawa2015,Louca2016}. 
In this data we lack access to the functional outputs of the community: we cannot measure how much nitrate is being dynamically consumed, only the steady state concentration. 
We nevertheless treat the environmental variables as a proxy for functional measurements, and train our networks to predict sample nitrate concentrations. 
Two groups suffice to accurately predict nitrate, and we find that the structure-function map is well-approximated by a linear model (Fig.~\ref{fig:validation}i-l and SI~Fig.~\ref{si_fig:fig3_method_comparison}). 
Because we lack a ground truth in this case, we confirm that our learned functional groups contain all strains identified by an alternative method (Ensemble Quotient Optimization; EQO) introduced in Ref.~\cite{Shan2023}. Like EQO, SCiFI produces a group consisting of ``complementary'' (anti-correlated) species, but in contrast to EQO it does so without having this feature built in  the objective function (SI~Fig.~\ref{si_fig:fig3_specspec_corrs}).

Finally, we consider soil microcosm experiments from Ref.~\cite{lee2024functional} which measure nitrate reduction dynamics together with the end-point phyla abundances under a range of pH perturbations. In contrast to the previous examples, function is a vector representing the concentration of nitrate over time.
Our method indicates that three functional groups, which in particular isolate Proteobacteria and Bacillota, are necessary to predict nitrate dynamics (Fig.~\ref{fig:validation}m-p).
This is consistent with previous work showing the key role of members of these phyla in governing denitrification in soils subjected to basic perturbations \cite{lee2024functional}.
These groups are sufficient to predict nitrate consumption only if the structure-function map is nonlinear (Fig.~\ref{fig:validation}o-p and SI Fig.~\ref{si_fig:fig3_method_comparison}).

\clearpage


\subsection{SCiFI distills groups of genes in the ocean metagenome}

\begin{figure}[tp]
    \centering
    \includegraphics[width=0.99\linewidth]{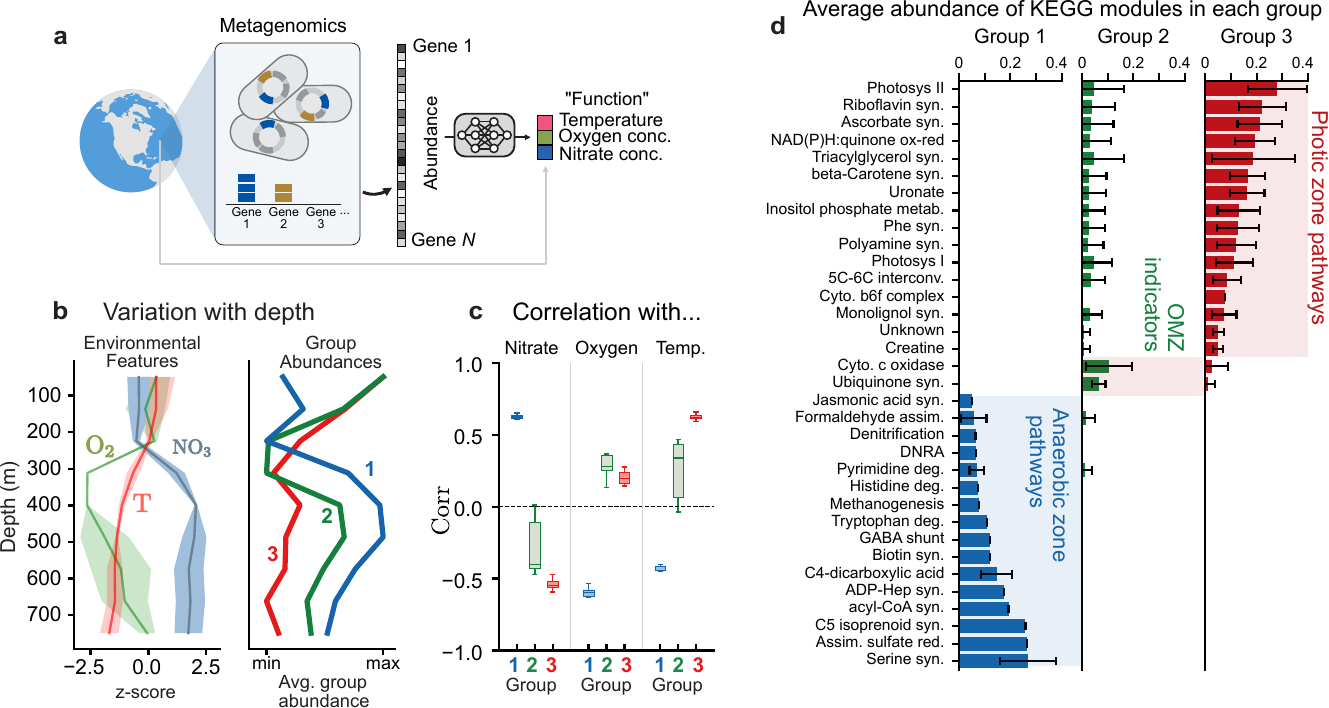}
    \caption{
    \textbf{Sparse functional gene groups reveal survival strategies in the ocean microbiome}
    \\
    (a) We use SCiFI to cluster gene modules using environmental parameters as a proxy for function.
    Structure in this dataset is quantified via shotgun metagenomics which quantifies the abundances of different genes across all genomes in the sample. 
    Environmental variables include both those related to community function (e.g. oxygen, nitrate concentrations) and to abiotic forcing (e.g. temperature).
    (b, left) Average nitrate concentration, oxygen concentration, and temperature as a function of depth. Quantities are normalized to have mean zero and standard deviation of one.
    (b, right) Average group abundance for the three groups identified with our algorithm, normalized by max/min.
    (c) Correlation of group abundances with each of the three environmental parameters used as target function during training: nitrate, oxygen, and temperature.
    (d) Allocation of several selected KO pathways to our learned groups. Bar heights denote the average abundance of each module assigned to each group. Average is taken across samples and across 12 test-train splits of the dataset; error bars show one standard deviation across test-train splits.
    }
    \label{fig:tara_metagenomics}
\end{figure}

Nothing about SCiFI limits its applicability to clustering species -- it just as readily clusters genes.
We now demonstrate this capability by finding relevant groups of genes present in microbial communities in the world's oceans.
The Tara Oceans dataset \cite{Sunagawa2015} provides measurements of gene abundances, which we consider aggregated at the level of KEGG modules \cite{kegg_2000,kanehisa_kegg_2025},  and environmental conditions at $\sim 100$ sites across the global ocean (Fig.~\ref{fig:tara_metagenomics}a). 
The large number of gene modules ($\sim$500) poses a challenge to interpreting the resulting groups because even if there are a small number of groups, they will be composed of a large number of genes.
Here, we employ SCiFI's optional gating procedure to achieve an approximate twenty-fold reduction in the number of relevant genes, which facilitates interpretation of their individual roles.

Similarly, nothing about SCiFI limits its applicability to clustering based on function alone -- it can also incorporate the role of abiotic forces. 
The Tara data includes abiotic factors known to structure marine microbiomes, such as temperature \cite{Louca2016}.
To account for the impact of this key abiotic factor in metabolic function, we ask SCiFI to predict temperature along with oxygen and nitrate concentrations.
Three groups suffice to make accurate predictions  (Fig.~\ref{fig:tara_metagenomics}b; SI Fig.~\ref{si_fig:tara_err_vs_nclust}). Two groups are strongly correlated to nitrate concentration (blue; group 1) and temperature (red; group 3) (Fig.~\ref{fig:tara_metagenomics}c), while the third (green; group 2) shows clear depletion at the transition to the oxygen minimum zone.

After the  sparsity promoting (gating) step, each group consists of only a small number of gene modules Fig.~\ref{fig:tara_metagenomics}d (also SI~Figs.~\ref{si_fig:tara_modules},~\ref{si_fig:fig4_metag_hparam_consistency}). 
This allows us to interpret how the enrichment of groups at certain depths (Fig.~\ref{fig:tara_metagenomics}b) corresponds to different survival strategies.
For example, group 1, which is abundant in deep waters, contains modules for respiration with alternative electron acceptors (nitrogen, sulfur). 
More interestingly, it also contains modules related to degradation of pyrimidine, histidine, and tryptophan.
This suggests that the bacteria residing in the nutrient poor conditions of the deep ocean survive by scavenging nucleotides and amino acids for energy and growth.
Group 3, in contrast, contains several gene modules related to protective compounds such as pigments (beta-Carotene), protective vitamins (ascorbate) and exopolysaccharide (uronate pathway), a key component of biofilms \cite{decho1990microbial,decho_microbial_2017}. 
These compounds might reflect adaptations to the high stress that surface bacteria experience at the surface due to UV radiation \cite{herndl_major_1993} or phage predation \cite{doolittle1995lytic}.

\subsection{SCiFI learns functional groups that are dynamical variables of mathematical models}

Up to now, we have interpreted the functional groups learned by SCiFI by looking at the phenotypes of group members, their phylogeny, or in the case of clustering metagenomes, the function of proteins those genes encode. However, this analysis does not reveal how the groups work together to drive community function since the structure-function map is encoded by a neural network that is not readily interpretable.
We now show that the functional groups learned by SCiFI can be directly used as variables in a fully interpretable mathematical model that relates groups to community function. 

As an example, we consider how nitrate utilization in soil microbiomes is impacted by short and long-term changes in soil pH, a key environmental variable \cite{bahram_structure_2018,lee2024functional}. 
Previous work has shown that a consumer-resource model predicts the dynamics of nitrate utilization (the resource) from a single effective biomass (the consumer) \cite{lee2024functional}.
The model parameters, including the size of the effective biomass, were inferred from the nitrate measurements alone. The approach revealed broad mechanisms of pH induced changes in function, but did not provide a principled way to identify the relevant bacteria from sequencing measurements.
Here we show that SCiFI solves this problem.

We train SCiFI to predict trajectories of nitrate concentration using a vector of 4395 amplicon sequence variant (ASV) abundances as input.
This data comes from experiments which measure the consumption of amended nitrate over time in soil samples collected across a natural pH gradient (Methods). ASV abundances are measured at the initial and final time points.
Samples are subjected to a range of pH perturbations and incubated with and without a growth inhibiting drug (chloramphenicol), yielding a total of $\sim$750 microcosms (Methods). 

We find that, with gating, two groups suffice to predict nitrate dynamics across the entire dataset (SI~Fig.~\ref{si_fig:soil_err_vs_nclust},~\ref{si_fig:soil_group_consistency}).
To capture how the groups learned by SCiFI collectively consume nitrate, we augment the consumer-resource model proposed in Ref.~\cite{lee2024functional}:
\begin{align}
    \dot x_i &= \gamma_ir_ix_i\frac{A}{A+K_A}\frac{C}{C+K_C}\,\,,\,\,i=1,2
    \\
    \dot A &= -(r_1x_1 + r_2x_2)\frac{A}{A+K_A}.
    \label{eq:consumer_resource}
\end{align}
The two biomasses $x_1$ and $x_2$ are readily identified as the learned functional groups which consume nitrate $A$  with yields $\gamma_i$ (1/mM) and at rates $r_i$ (mM/day) which vary with pH. Biomass growth is limited by the availability of nitrate and carbon $C$; the dynamical equation for the latter is not shown (see Methods). 
The measured group abundances $x_i(0)$ (initial) and $x_i(T)$ (final), together with the measured nitrate dynamics, serve as constraints to fit the parameters $\gamma_i$ and $r_i$ (Methods).

\begin{figure}[t] 
    \centering
    \includegraphics[width=0.5\textwidth]{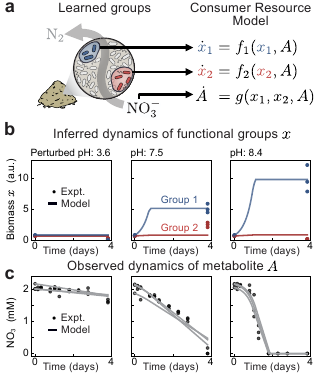}
    \caption{\textbf{Learned functional groups are the relevant variables of minimal dynamical models}
    (a, left) In soil microcosm experiments quantifying nitrate utilization \cite{lee2024functional}, SCiFI finds that only two groups are necessary.
    (right) These group abundances, together with nitrate measurements, are described by a simple consumer resource model.
    (b) Evolution of group abundances $x_1$ (blue) and $x_2$ (red) in time. Measurements are taken at $t=0$ and $t=96$h after incubation, with three replicates per sample (each point is one replicate).  Colored lines denote inferred dynamics from the consumer resource model, where we take the average across replicates Eq.~\eqref{eq:consumer_resource}.
    (c) Evolution of nitrate concentration over time. Concentrations are measured at ten time points. Each gray line is the consumer-resource prediction for one replicate.
    }
    \label{fig:soil_CR_dynamics}
\end{figure}

The resulting model accurately predicts both nitrate dynamics and the observed biomass abundances across different pH levels (Fig.~\ref{fig:soil_CR_dynamics}b-c). Notably, biomass abundances are determined by inferring their unseen dynamics (Fig.~\ref{fig:soil_CR_dynamics}b). We evaluated the importance of our groups by comparing to a null model which uses randomly selected groupings (SI Fig.~\ref{si_fig:soil_null_models} and Methods).
Randomization of either group significantly degrades nitrate predictions which shows that the groups found by SCiFI are uniquely informative of the nitrate consumption dynamics.

\clearpage
\newpage

\begin{figure}[tp]
    \centering
    \includegraphics[width=0.9\textwidth]{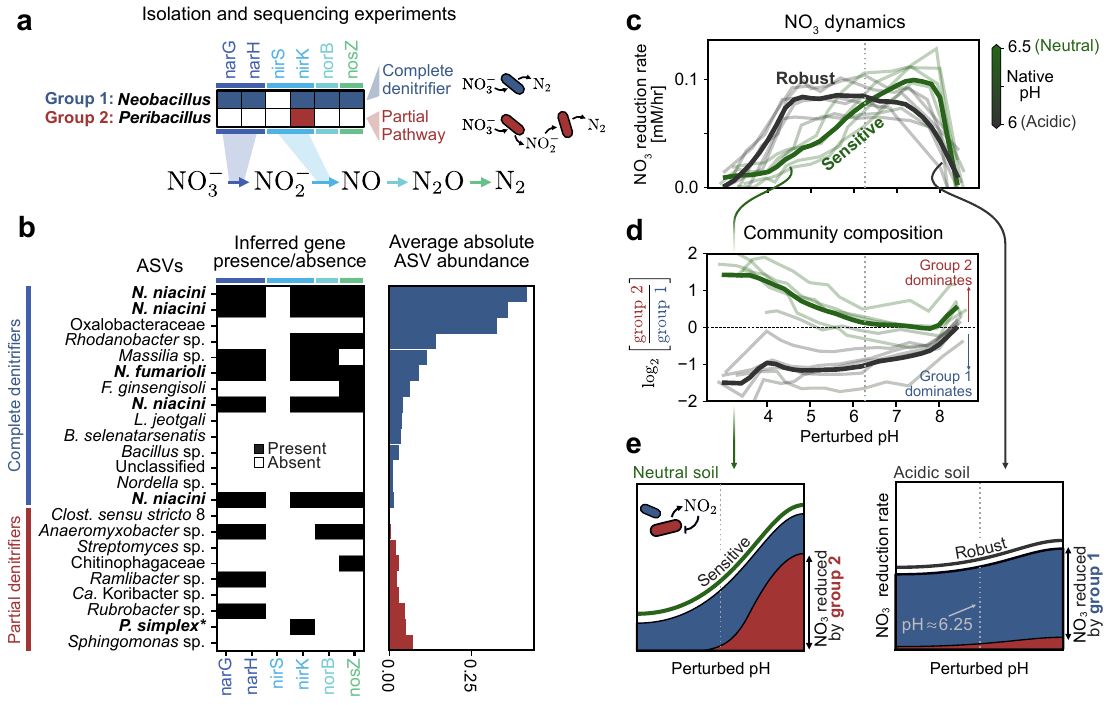}
    \caption{
    \textbf{Functional groups reveal mechanistic underpinnings of pH sensitivity in soil communities}
    (a) Whole genome sequencing of two isolates from the groups shown in (b) with the denitrification relevant enzymes they possess. \textit{Neobacillus} has enzymes to perform all steps in the cascade (bottom), while \textit{Peribacillus} can only perform the second step. 
    (b, left) 
    Denitrification gene presence (black) and absence (white) for individual ASVs in group 1 and group 2. Gene presence for strains in the genera isolated in (a) is inferred from sequencing data, for all other strains we use PICRUSt2 \cite{douglas_picrust2_2020}.
    (right) Average end-point abundance of the selected ASVs, where average is taken across all samples not subjected to growth inhibition via chloramphenicol.
    (c) Nitrate reduction rate ($y$-axis) of soils varies depending on perturbed pH ($x$-axis). Each faint line corresponds to one soil sample treated with chloramphenicol (non-treated samples are shown in SI Fig.~\ref{si_fig:no3_reduction_chlplus}). Solid lines correspond to the average when grouped by pH; acidic (gray) soils are those with native pH below 6.25, neutral (green) soils are those with pH greater than 6.25. 
    In all panels in (c-e), the gray dashed vertical line denotes a perturbed pH of 6.25.
    (d) Variance of the community composition, as measured by log ratio between group 1 and group 2, as a function of perturbed pH. A value greater than 0 indicates that group 2 dominates, and a value less than zero indicates group 1 dominates. As in (c), faint lines indicate individual soils and solid lines are the average across acidic (gray) and neutral (green) soils.
    (e) Simple picture showing how the observed dependence of nitrate reduction on pH arises due to our two groups. 
    (left) Neutral soils have, at neutral pH, a large contribution to nitrate reduction from both group 1 (blue) and group 2 (red). As pH is decreased, the contribution due to group 2 disappears because of nitrite toxicity (inset). 
    (right) In acidic soils, only group 1 is present. Because it is a complete denitrifier, it is not sensitive to the buildup of nitrite and hence robustly reduces nitrate across all pH levels.
    }
    \label{fig:soil_genomics}
\end{figure}

\subsection{SCiFI guides targeted experiments that reveal biological mechanisms}

The two functional groups we find exhibit distinct dynamical behaviors (Fig.~\ref{fig:soil_CR_dynamics}b).
What is the biological explanation for this difference? SCiFI by itself does not answer this question. However, it does perform the first step of an integrated pipeline that combines machine learning with experiments to reveal underlying biological mechanisms. Concretely, our approach exploits the sparse groups learned by SCiFI to guide targeted experiments on representative species.



For nitrate reduction in soil, we seek a genetic explanation.
This choice is motivated by the fact that genes encode nitrate utilization traits of individual species \cite{gowda_genomic_2022}. We isolated a dominant strain from each group (\textit{Neobacillus fumarioli} from Group 1; \textit{Peribacillus simplex} from Group 2) and sequenced their genomes (Methods). These strains were selected because they were among the most abundant in their respective groups. 
Because these strains differ only at the genus level, the distinction between them would not have been discovered without function-informed clustering.

Whole genome sequencing of these isolates revealed that, despite their taxonomic proximity, these strains differ in the denitrification enzymes they harbor (Fig.~\ref{fig:soil_genomics}a). While the \textit{Neobacillus} strain (blue) possesses a complete set of genes to reduce nitrate to dinitrogen, the \textit{Peribacillus} strain (red) can only perform the intermediate step of reducing nitrite to nitric oxide. 
In Figure~\ref{fig:soil_genomics}b we show the inferred genomes of all group members (PiCRUSt2, \cite{douglas_picrust2_2020}) sorted by their average absolute abundance (right).
We observe the same pattern: the most abundant Group 1 strains (blue) often have a full set of genes required to convert nitrate to dinitrogen, while Group 2 strains perform only partial steps in the pathway (Fig.~\ref{fig:soil_genomics}b).

This finding allows us to explain how the nitrate reduction capacity of a community changes in response to pH perturbations.
In growth-inhibited conditions we observe that acidic soils (pH$\lesssim 6$; gray in Fig.~\ref{fig:soil_genomics}c) show robust nitrate reduction which is constant over a range of pH perturbations, while neutral soils (pH$\gtrsim 6.5$; green in Fig.~\ref{fig:soil_genomics}c) are more pH-sensitive.
These two soil types differ also in their composition: Group 1 dominates acidic soils, while Group 2 dominates neutral soils (Fig.~\ref{fig:soil_genomics}d).

Taken together, these observations suggest the following mechanism to explain community sensitivity to pH perturbations. 
Neutral soils are characterized by a large population of Group 2, which reduces nitrate by splitting the pathway among several different constituent strains. Such cross-feeding communities are susceptible to nitrite toxicity at low pH \cite{goldschmidt_metabolite_2018}, which slows the overall utilization of nitrate (Fig~\ref{fig:soil_genomics}e, left). Acidic soils, by contrast, are dominated by Group 1 which is expected to fully reduce nitrate to dinitrogen and relieve nitrite toxicity, resulting in robustness to pH perturbations (Fig.~\ref{fig:soil_genomics}e, right). 
Both behaviors persist in growth conditions (SI Fig.~\ref{si_fig:no3_reduction_chlplus}). 
Our integrated pipeline reveals how collective function depends on the differential response of individual groups to perturbations.

\section{Discussion}

It has been known since the Babylonians first made sourdough that communities of bacteria perform important macroscopic functions. Only in the last century was it  discovered that these communities are composed of groups, or guilds, of microbes with similar metabolic strategies. The challenge for microbiologists ever since has been to identify these strategies and disentangle their contributions to collective metabolism. 
A common approach to identify groups and their strategies uses enrichment cultures. 
For example, incubating a soil community in nitrate-rich, anoxic conditions will enrich microbes that perform nitrate reduction. 
Genomics provides an alternative route to identify groups in natural settings based on the presence or absence of genes that encode for functionally-relevant enzymes. 
Neither of these approaches, however, quantitatively identifies a microbe's contribution to community function. Enrichment cultures necessarily remove microbes from their natural context and hence do not capture how a microbe behaves when embedded in a complex community, while genomics does not consider metabolite fluxes explicitly.

Our approach is not subject to these shortcomings because it learns functional groups directly from data of natural communities. 
This success hinges on the fact that SCiFI is function-informed and nonlinear. 
Because it is function-informed, we do not need experiments to be fine-tuned to elicit the growth of one particular functional group. Instead, groups can be identified \textit{post hoc} from any experiment that generates correlations between abundances and function. Nonlinearity allows SCiFI to capture interactions between groups, such as those between \textit{A. caccae} and pH buffering strains that modulate modes of butyrate production (Fig.~\ref{fig:validation}a-d).

SCiFI can dramatically simplify the process of characterizing the structure of the community and its relationship to function.
Because it is function-informed, SCiFI can leverage a single experiment that measures multiple functions to learn multiple groupings from the same abundance data. 
This could dramatically reduce the number of experiments needed to characterize the structure of a community.

Equation learning pipelines that build on SCiFI can explain the relationship between structure and function. Here, we used SCiFI to extract the relevant variables for a consumer resource model whose mathematical form was chosen based on intuition grounded in previous work \cite{lee2024functional}.
However, if we replace the neural network structure-function map with a sparse-regression model \cite{brunton_discovering_2016} or a NeuralODE \cite{neuralode_chen_2018}, the form of the mathematical model itself can be inferred directly from data with little  \textit{a priori} knowledge.

Complex structure-function relationships are not unique to microbiomes.
Our integrated pipeline can find low-dimensional descriptions of structure in other contexts with no change to the core algorithm. 
The underlying assumption behind SCiFI is that group members (of which bacterial species are but an example) are indistinguishable and may meaningfully be aggregated by summation. 
Any problem that has this feature may be tackled with our framework.
We suggest two cases where this assumption is likely to be fulfilled. 
First, SCiFI may be applied to neural firing rates to learn the low-dimensional ``neural manifold'' underlying motor functions \cite{gallego_neural_2017}. 
For example, a simple aggregation of (preprocessed) neural activity has been shown to predict the direction of rhesus monkey arm movement from single-neuron data alone \cite{georgopoulos_neuronal_1986}. SCiFI could automate this aggregation procedure and extend it to non-linear structure-function mapping.
As a second example, SCiFI may be used to understand the structure-function relationship between T-cell receptor sequence and immune response. 
The immune response to a particular antigen is determined by the aggregated activity of all T-cells which recognize it \cite{immuno_buchholz_disparate_2013}. SCiFI could learn functional groups from experiments which challenge a T-cell community with a complex pathogen (e.g. a whole virus) and measure their collective response (e.g. total interferon $\gamma$ production).

\section{Limitations of the study}
SCiFI identifies groups based on a statistical relationship between abundances and function. However, it is not guaranteed that these groups are actually generators of function. This is a general limitation of machine learning methods that can be ultimately traced to the fact that correlation is not causation. 
While in every dataset we considered SCiFI learned biologically relevant groups, this outcome ultimately needs to be independently checked either with mathematical modeling or further experiments.

As a data-driven approach, SCiFI inevitably inherits the limitations of the dataset. 
Noise can hinder SCiFI's ability to find groups, especially groups that only weakly impact function. 
For example, in a simulated linear degradation cascade where function is defined as the concentration of the final product, groups that are far upstream are recovered imperfectly even though they are biologically necessary (SI Fig.~\ref{si_fig:linear_degradation_chain}).

Finally, as described above, the grouping sought by SCiFI inherently assumes that species within a group are indistinguishable. 
While this limits the class of problems to which it may be applied, it is also crucial for the interpretability of our method. 
SCiFI may be adapted to allow for distinguishable group members in a way analogous to how we introduced the gating to enforce sparsity. However, this extension may come at the cost of reduced    interpretability.

\end{linenumbers}

\setcounter{figure}{0}
\renewcommand{\thefigure}{S\arabic{figure}}
\renewcommand{\figurename}{Supplementary Figure}

\clearpage
\section*{Methods}
\subsection*{Neural network details}
Here we describe the architecture and training protocol of SCiFI in detail. 
Let us denote the abundance (row) vector for one sample by $x\in\mathbb{R}^n$ and the corresponding function (row) vector by $f\in\mathbb{R}^m$. 
To make a prediction of the function from the abundance data, SCiFI first computes group abundances $g\in\mathbb{R}^{N_{\text{clust}}}$ via the clustering matrix $C\in\{0,1\}^{n\times N_{\text{clust}}}$, so that $g=xC$. (We discuss how the clustering matrix $C$ is generated below.)
From the group abundances, we predict the function with a neural network parameterized by parameters $\theta$: $\hat{f}=\text{NN}_{\theta}(g)$. 
From this, we compute the mean squared error as our loss function
\begin{align}
    \mathcal{L}(C, \theta)=\frac{1}{B}\sum_{i=1}^{B}\lVert f_i-\hat{f}_i(x_i;C,\theta)\rVert^2
\end{align}
where the sum is taken over samples in our batch $B$. In the case of minibatch gradient descent this is a random subset of the data, but due to the small number of samples in our datasets considered here we do not need to perform minibatching.

\subsubsection{Gated SCiFI for sparse groups}
The gating procedure described in the main text is easily incorporated as a step prior to the grouping. The gate is represented by a vector $\gamma\in(0,1)^n$, and is applied to get the gated abundances $x^{\text{g}}=\gamma\,\odot\,x$, where $\odot$ denotes element-wise multiplication. The gate is computed from a real-valued parameter $\ell_i\in\mathbb{R}$ via $\gamma_i=\text{sigmoid}(\ell_i)$ which allows us to learn it with standard gradient descent.
From there, group abundances are computed as $g=x^{\text{g}}C$ and the following steps are unchanged. The loss function includes a term to enforce sparsity in the gate, 
\begin{align}
    \mathcal{L}(C, \gamma,\theta)=\frac{1}{B}\sum_{i=1}^{B}\lVert f_i-\hat{f}_i(x_i;C,\gamma,\theta)\rVert_2^2 + \beta\lVert\gamma\rVert_1
\end{align}
where here we distinguish between the 2-norm $\lVert\cdot\rVert_2$ and 1-norm $\lVert\cdot\rVert_1$. We introduce a new hyperparameter $\beta$ which controls how much we want to enforce sparsity; a large $\beta$ leads to very sparse solutions, but typically there is a trade-off between sparsity and predictive accuracy. 

\subsubsection{Gumbel softmax for clustering}
Here we explain how the Gumbel softmax trick may be used for clustering.
The clustering problem is one of assigning a cluster index to every entry in our input vector. If $x\in\mathbb{R}^n$ is our vector of abundances, then the clustering is represented as a vector $\hat{c}\in\{1,...,N_\text{clust}\}^n$ where $N_{\text{clust}}$ is the number of clusters, for example $\hat{c}=(1,2,1,1,3,2)$ represents a clustering of 6 species into 3 clusters.
Instead of representing the cluster identity as a (categorical) integer, we may equivalently represent it by a one-hot vector. The clustering is then given by a matrix $C\in\mathbb{R}^{n\times N_{\text{clust}}}$. The above example would be 
\begin{align*}
    C=\begin{pmatrix}
        1 & 0 & 0 \\
        0 & 1 & 0 \\
        1 & 0 & 0 \\
        1 & 0 & 0 \\
        0 & 0 & 1 \\
        0 & 1 & 0 
    \end{pmatrix}.
\end{align*}
Each row is the cluster assignment of a single species.
The question is how to encode this matrix with continuous variables so that we may perform gradient descent.

\begin{figure}[t]
    \centering
    \includegraphics[width=0.5\linewidth]{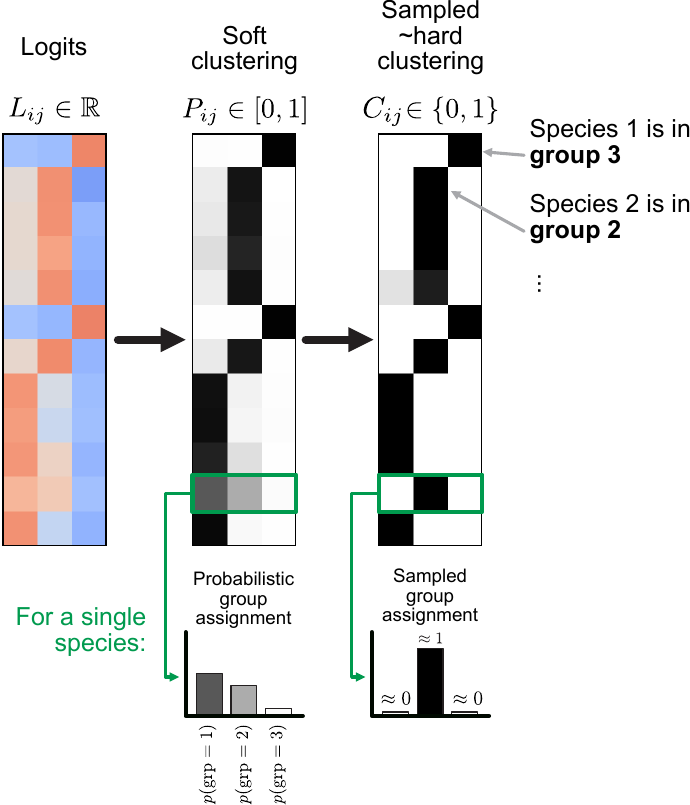}
    \caption{\textbf{Illustration of the Gumbel softmax trick for generating a clustering matrix}
    The central matrix $P$ denotes the probabilistic clustering matrix. It is derived (via softmax) from the matrix $L$ to the left, the real-valued matrix of logits which are updated by gradient descent. 
    Each row of $P$ describes the probabilistic group assignment of a single species to different groups.
    From $P$, an approximately \textit{hard} clustering matrix $C$ is sampled using the Gumbel softmax trick described in Methods. Each row is approximately a one-hot vector that describes the group identity of the species.
    }
    \label{si_fig:gumbel_demo}
\end{figure}

This is achieved with the Gumbel softmax trick \cite{gumbel_maddison_2014, gumbel_maddison_2017,gumbel_jang_2017}, a method for differentiably sampling categorical variables.
For us, the categorical variable is the cluster identity (e.g., a single entry in the vector $\hat{c}$). 
This sampling scheme will essentially allow us to ``try'' many different possible clusterings $\hat{c}$ during training.
To do this, each one-hot encoding of cluster identity (one row in $C$) is replaced by a soft assignment $\pi$ of length $N_{\text{clust}}$. 
Instead of being a one-hot vector, $\pi$ represents the \textit{probability} of being assigned to each cluster. For example, $\pi=(0.08, 0.90, 0.02)$ says that there is an 8\% chance of being clustered into group 1, a 90\% chance of being clustered into group 2, etc. While we hope that this converges to a hard clustering during training so that each species is deterministically assigned to one group (i.e. probability of assignment becomes a delta function on group $k$: $\pi=(0,...,\underset{k}{1},...,0)$), in practice this is not always the case. This represents real uncertainty about which group a species is assigned to.
The distributions for all species are represented by a matrix $P\in[0,1]^{n\times N_{\text{clust}}}$. Each row is distribution $\pi_i$ for species $i$ and sums to one.

\begin{figure}[t]
    \centering
    \includegraphics[width=0.99\linewidth]{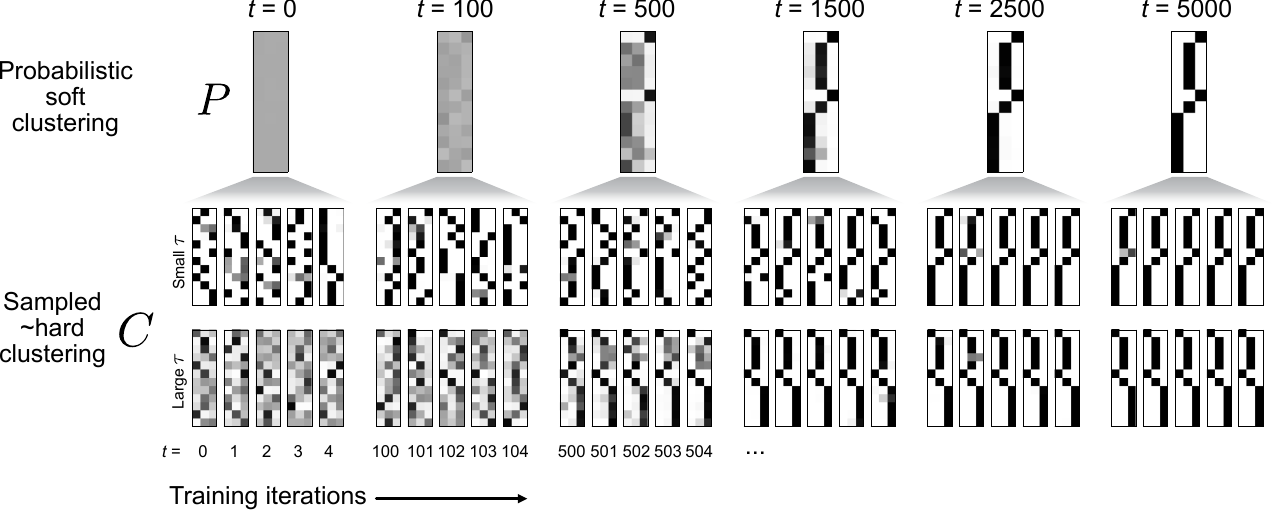}
    \caption{Illustration of $P$ and $C$ matrices during training on a toy dataset. (Top) The probabilistic clustering $P$ evolves slowly during training. (Bottom two rows) At each training step, an approximately hard clustering matrix $C$ is sampled. For small $\tau$ (middle) this approximation is closer.}
    \label{si_fig:gumbel_demo_in_time}
\end{figure}

\begin{figure}[tp]
    \centering
    \includegraphics[width=0.9\linewidth]{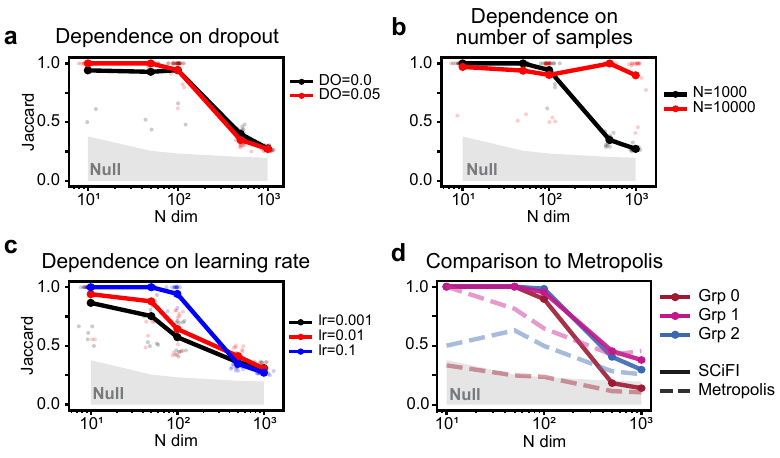}
    \caption{\textbf{Training hyperparameters impacting performance on synthetic data.}
    We explore how SCiFI's ability to recover the correct groups depends on critical training parameters and on the dimension of the input. 
    The synthetic data follows the structure discussed in Methods and used in Fig.~\ref{fig:the_model}. The basic group structure is $\hat{c}=(1,2,2,2,3,3,3,3,3,3)$ which assigns 10 species to 3 clusters. We vary the dimensionality of the system while keeping the group number fixed by simply repeating $\hat{c}$ for an integer number of repetitions (e.g. 10 repetitions would correspond to a dimensionality of 100). 
    The neural network used to model the structure function map is small, consisting of 2 layers of 48 neurons each. This small network is sufficient because the function is piecewise linear.
    In all cases we measure group recovery with the Jaccard index (see Methods).
    (a) Variation of group recovery as a function of dimensionality for different values of dropout. 
    Every faint dot corresponds to one model of an ensemble ($N_{\text{ens}}=15$), slightly jiggled in the $x$-direction for visualization. The solid dots with connecting line are the average Jaccard index across all ensemble members.
    The number of samples used for training is 1000 and we see performance start to decrease at $N_{\text{dim}}=100$.
    Dropout makes negligible difference to the group recovery, but it significantly improves generalization (not shown). 
    (b) Same as in (a) but varying the number of training samples. Performance increases significantly when the number of samples is increased. This suggests that given enough data, SCiFI should find the correct groups.
    (c) Same as in (c) but varying the learning rate during training. In general, learning rate can significantly impact performance. Here we see that using a larger learning rate improves group recovery. However, too large and the networks will fail to train (not shown).
    (d) Comparison of SCiFI group recovery to the Metropolis Hastings algorithm from Ref.~\cite{Zhao2024}. 
    Here we show the recovery of each group individually. SCiFI generically outperforms the Metropolis Hastings algorithm. 
    Here we also see that the group recovery differs across groups. This suggests that the decrease in performance at $ N_{\text{dim}}\approx 100$ is due to the particular structure-function map used. 
    }
    \label{si_fig:synthetic_data_benchmarking}
\end{figure}

\begin{figure}[tp]
    \centering
    \includegraphics[width=0.99\linewidth]{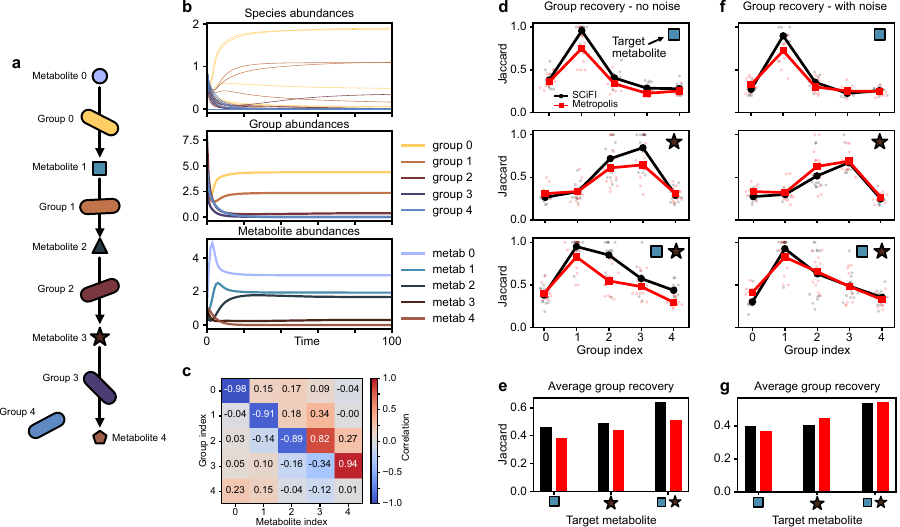}
    \caption{\textbf{SCiFI applied to a linear degradation chain.}
    (a) Here we evaluate SCiFI on a simulated community of species which participate in a linear degradation chain of a metabolite as in Ref.~\cite{Zhao2024}.
    The first four groups participate in the chain, the last group does not. 
    (b) Evolution in time of the species abundances (top), grouped abundances (middle) and metabolite abundances (bottom). 
    (c) Correlations between end-point group abundances and metabolite abundances.
    (d) Group recovery as measured by the Jaccard index for each group. Red shows the results of the Metropolis-Hastings algorithm from Ref.~\cite{Zhao2024}, black shows SCiFI. The icon in the upper right indicates what function (metabolite abundance) was used to train SCiFI. The top row uses metabolite 1, middle uses metabolite 3, and bottom uses both. 
    (e) Average group recovery across all groups, separated by target metabolite algorithm (SCiFI - black, Metropolis - red) . Group recovery improves when using multiple intermediate metabolites (right columns).
    (f-g) Same as in (d-e) but with added 10\% noise on the metabolites concentrations. We see that the performance gap between SCiFI and Metropolis decreases. This is unsurprising because Ref.~\cite{Zhao2024} shows that the Metropolis algorithm works well in this setting.
    }
    \label{si_fig:linear_degradation_chain}
\end{figure}

\begin{figure}[tp]
    \centering
    \includegraphics[width=0.3\linewidth]{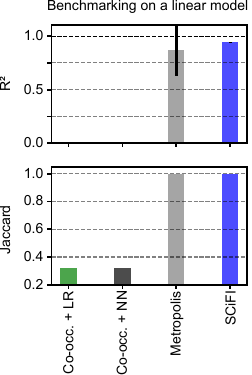}
    \caption{\textbf{Benchmarking on a linear model.}
    This plot compares the same models as in Fig.~\ref{fig:the_model}d-e, but when the true structure-function map is linear.
    In this case we see that Model III (the Metropolis-Hastings algorithm from Ref.~\cite{Zhao2024}) recovers the groups correctly.
    The co-occurrence models perform poorly because in this simulation there is no built-in correlation structure among different species. 
    }
    \label{si_fig:benchmark_linear_SF_map}
\end{figure}

\begin{figure}
    \centering
    \includegraphics[width=0.9\linewidth]{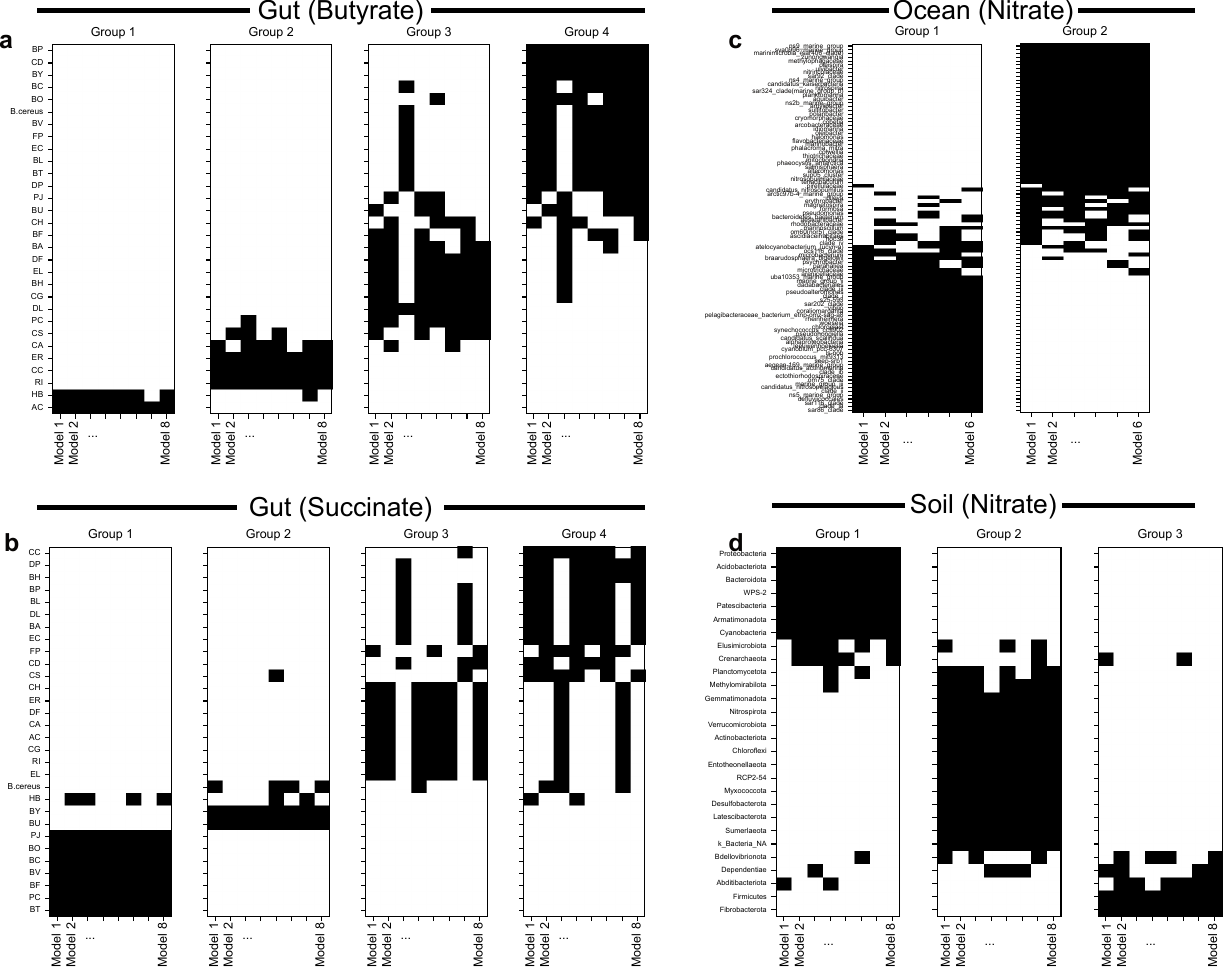}
    \caption{\textbf{Consistency of groups across test-train splits.}
    For each of the datasets in Fig.~\ref{fig:validation} we illustrate the consistency of the learned groups across the different test/train splits used for cross validation. 
    As a reminder, we train an ensemble of models with different test train splits (Methods). 
    (a) Each matrix (left to right) corresponds to one group. Each column of the matrices in this figure corresponds to one model. Each row corresponds to one species. Black means the species is included in the corresponding group, white means it is not. 
    For example, AC is always in Group 1, and CC is always in Group 2.
    Because the indices of the groups are arbitrary, we must reorder them so they are consistent across models. We do this by ordering the groups by their correlation with the target function. This process is imperfect as can be seen in Group 3 and Group 4, where it is clear that the assignments of the groups have been switched but the composition is unchanged. 
    (b) Corresponding plots for the gut data when trained to predict succinate using four groups (note that only two groups are used in Fig.~\ref{fig:validation}g-h).
    (c) Corresponding plots for the ocean data when predicting nitrate.
    (c) Corresponding plots for the soil data when predicting nitrate dynamics.
    }
    \label{si_fig:fig3_CV_consistency}
\end{figure}

\begin{figure}
    \centering
    \includegraphics[width=0.99\linewidth]{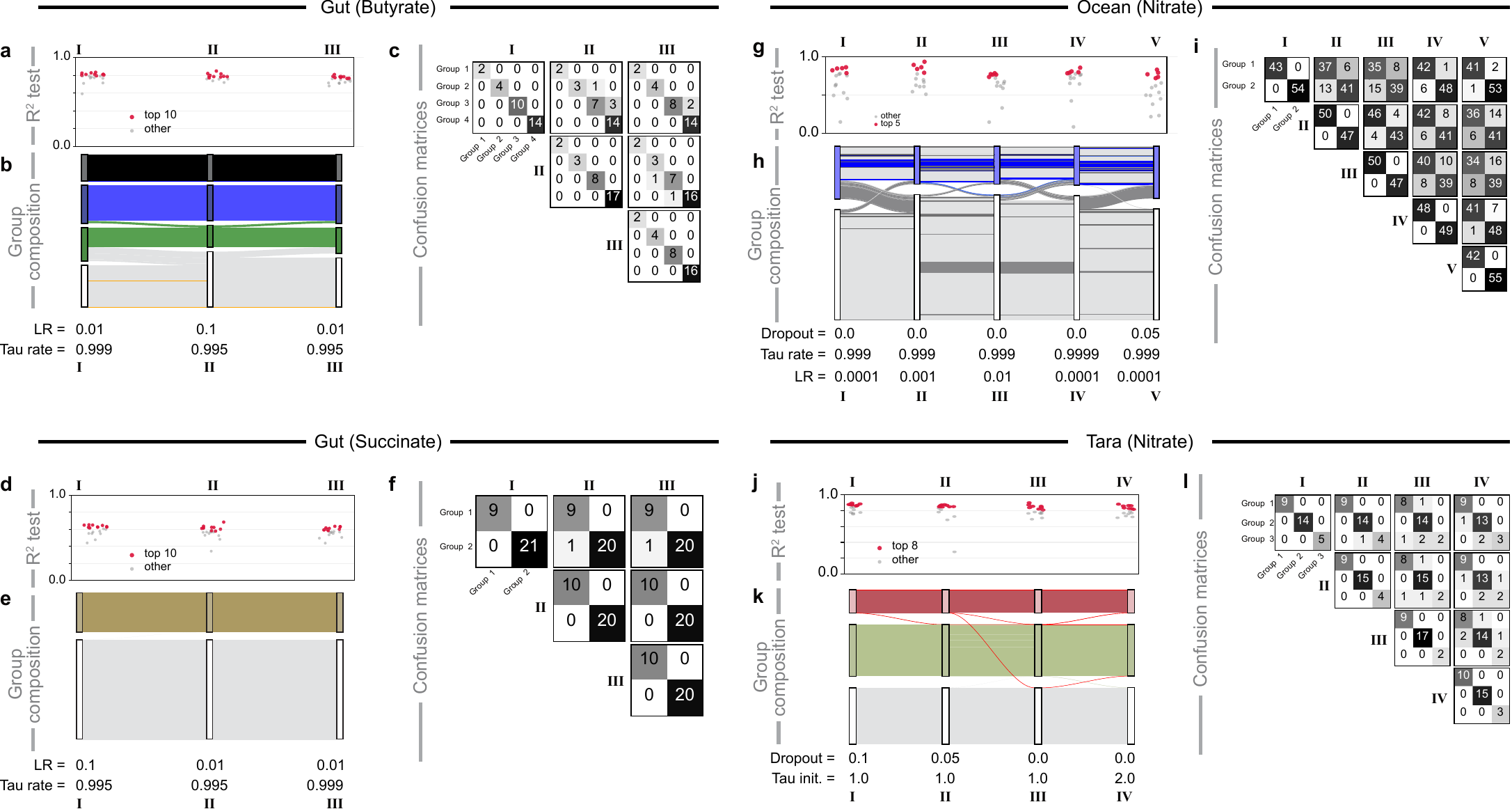}
    \caption{\textbf{Consistency of groups across hyperparameters.}
    For each of the datasets in Fig.~\ref{fig:validation} we evaluate the sensitivity of the learned groups to hyperparameters used for model training (Methods).
    For each of these datasets we show three plots which follow the same structure.
    (a) $R^2$ values for each of the different hyperparameter sets, enumerated with Roman numerals. Small dots correspond to single models in the ensemble. The top models highlighted in red are used to determine the consensus grouping (Methods).
    (b) Consensus group composition resulting for each hyperparameter set. 
    Each set of nodes (rectangles) in the flow diagram are aligned vertically with the corresponding hyperparameter set from (a). Each bar between nodes represents one species; its thickness corresponds to its average abundance. While the grouping is very stable for the synthetic gut data, it is less so for the ocean data (panels g-i).
    The hyperparameters corresponding to each set (I-III) are shown at the bottom of (b).
    (c) Confusion matrices to compare the group assignments across hyperparameter sets. Each matrix is a comparison between one set of hyperparameters (Roman numerals).
    Entry $i,j$ in the matrix is the number of species that are assigned to group $i$ for one hyperparameter set and to group $j$ for another hyperparameter set.
    (d-f) Corresponding plots for the gut data with succinate concentration as the target function.
    (g-i) Corresponding plots for Tara oceans data with nitrate concentration as the target function. Blue species which were identified by the EQO method from Ref.~\cite{Shan2023} are highlighted in blue.
    In this data (as opposed to the synthetic gut data) there is more variation in the relevant (top) group which reflects the fact that the data is both more complex (higher dimensionality) and less sampled. Species colored in dark gray are ones that are sometimes, but not always, in the relevant group.
    (j-l) Corresponding plots for the soil data with nitrate dynamics as the target function.
    }
    \label{si_fig:fig3_hparam_consistency}
\end{figure}

\begin{figure}
    \centering
    \includegraphics[width=0.99\linewidth]{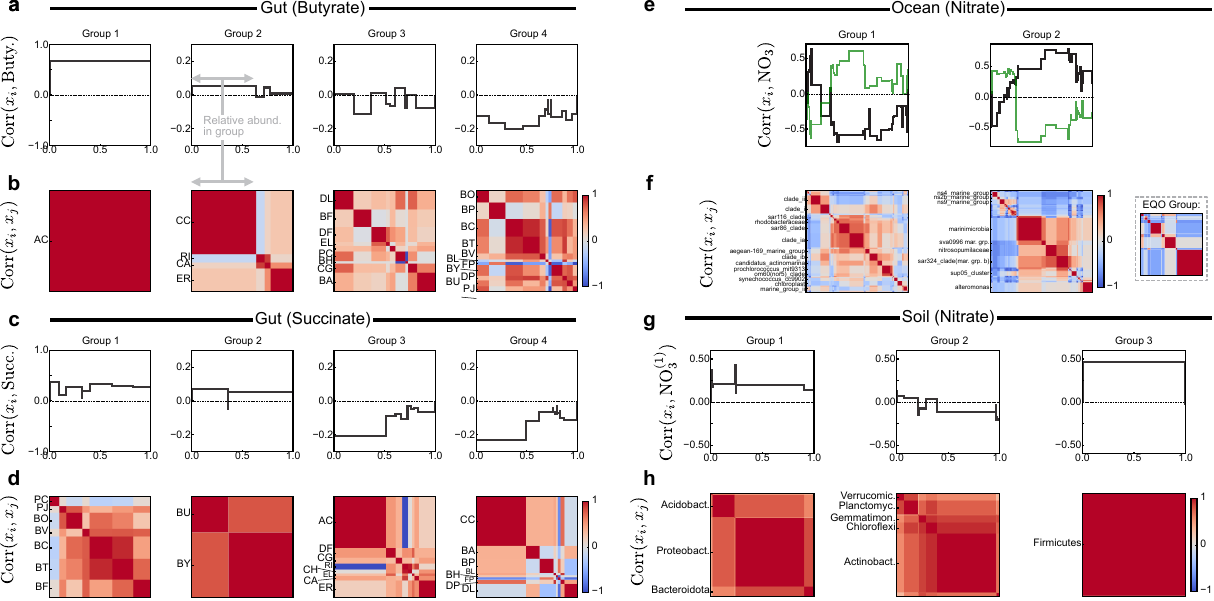}
    \caption{\textbf{Species-species and species-function correlations.}
    Here we investigate the composition of the groups in each of the datasets in Fig.~\ref{fig:validation}.
    For each dataset we visualize the correlations between individual species and the target function (top row, line plot) and species-species correlations (bottom row, image plot). 
    (a) For each group (column) we plot the correlation of the constituent species with butyrate. The order of the species is the same as in (b) and they are aligned visually. The width of each species corresponds to its relative abundance in the group. 
    Each block in the images in $i,j$ in (b) show the correlation between species $i$ and species $j$ (red is positive, blue is negative).
    For example, \textit{AC} completely dominates Group 1 and has a correlation of $\sim$0.7 with butyrate. 
    Group 2 is composed of four species. CC and CA are slightly positively correlated with butyrate, while ER is not correlated (despite being a direct producer). CC has the largest relative abundance in the group by far.
    (c-d) Same plots for the synthetic gut data when succinate is the target function. We observe two groups that are positively correlated with succinate (Group 1 and 2), while the others are negatively correlated.
    (e-f) Plots for the ocean data in Fig.~\ref{fig:validation}i-l. Group 2 is the group that contains all species identified by the EQO method from Ref.~\cite{Shan2023}.
    (e) In addition to showing how each species is correlated with nitrate (black) we show how they are correlated with oxygen (green). Typically these correlations are opposite sign, except for several species in Group 2. This suggests a similar group composition as identified by EQO \cite{Shan2023} (shown to the right), where species with different, ``complementary'' strategies are grouped to form a functional group that is informative across all environmental conditions. In this case, high and low oxygen concentration.
    (g-h) Plots for the soil data in Fig.~\ref{fig:validation}m-p.
    }
    \label{si_fig:fig3_specspec_corrs}
\end{figure}

\begin{figure}[tp]
    \centering
    \includegraphics[width=0.5\linewidth]{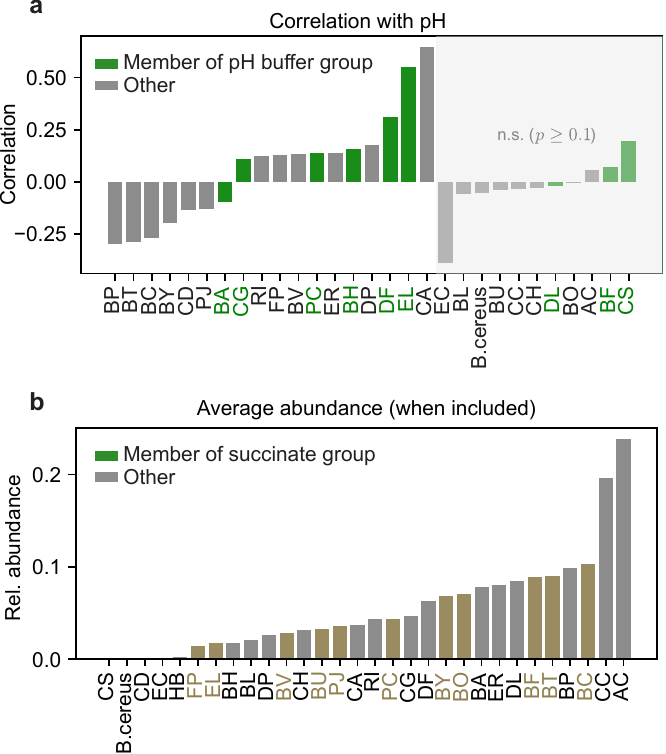}
    \caption{\textbf{Gut species abundances and correlations with pH.}
    (a) Correlation of every species with pH in samples where that species was included. Samples with $p$ value greater than $10^{-1}$ are on the right. Species in the ``pH buffer group'' are highlighted in green.
    (b) Average abundance of species across the dataset (not stratified by community richness). Species in the relevant succinate group are highlighted in brown.}
    \label{si_fig:clark_correlations_and_avg_abd}
\end{figure}

\begin{figure}[tp]
    \centering
    \includegraphics[width=0.6\linewidth]{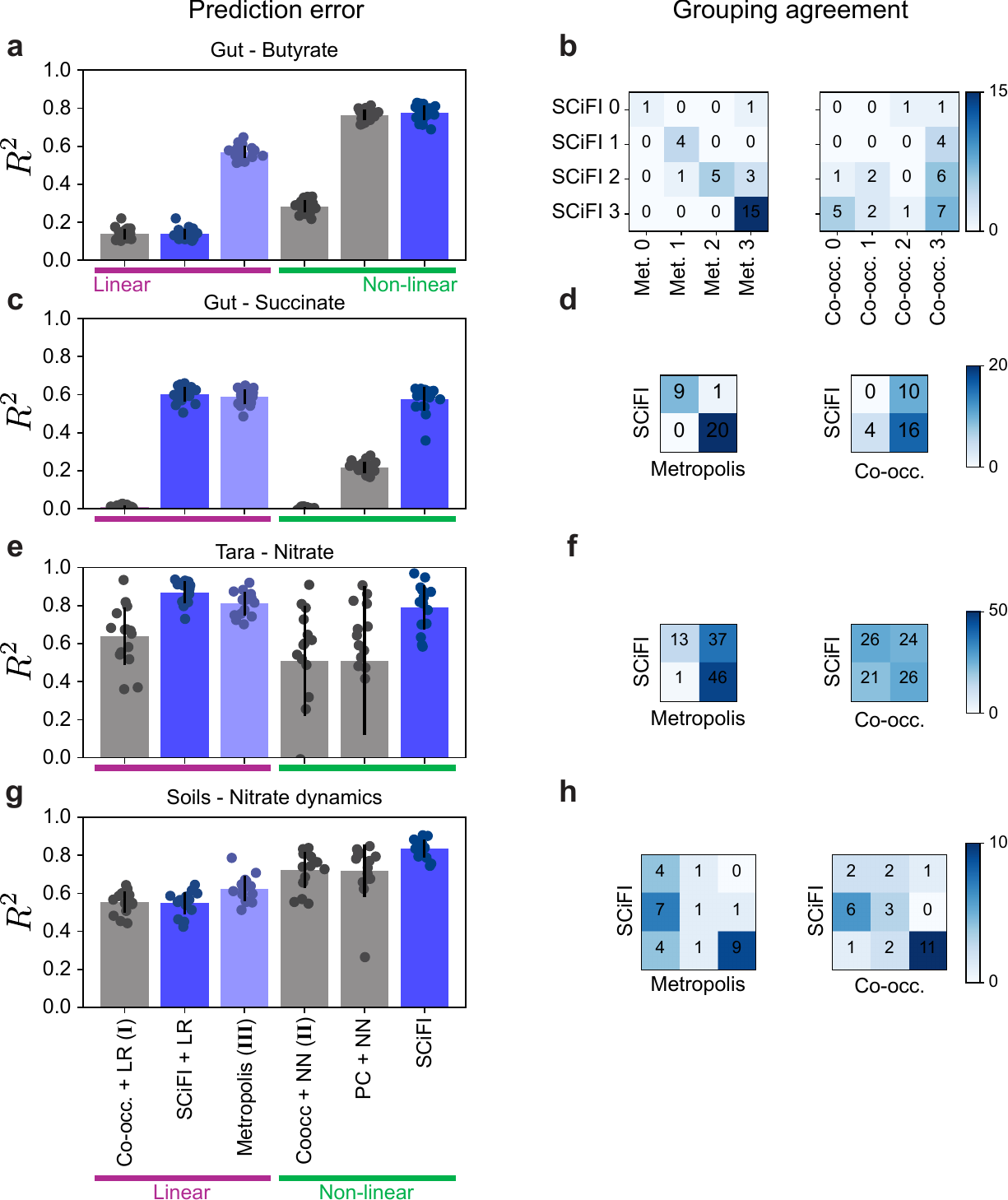}
    \caption{\textbf{Method comparison.}
    Each row corresponds to one row (dataset) in Fig.~\ref{fig:validation}.
    We compare SCiFI to five other approaches. For the clustering step, we consider either a co-occurrence network approach (Co-occ.), the clustering learned by SCiFI, or a linear function-informed clustering method that relies on a Metropolis-Hastings algorithm \cite{Zhao2024}. The co-occurrence clusters are either used to train a linear regression model (LR), corresponding to a Model I-type algorithm discussed the main text and Fig.~\ref{fig:the_model}, or to train a neural network (NN), corresponding to a Model II-type algorithm.
    In addition, we also consider a model which performs dimensionality reduction with principal component analysis (PCA). 
    Left shows the $R^2$ of these six different models. The first three have linear structure-function maps, the last three are nonlinear. 
    The bars show the average $R^2$ across the entire ensemble of models ($N_{\text{ens}}=$12-20 depending on dataset), with individual points showing the $R^2$ for single models.
    Blue bars are function-informed.
    Dark blue denotes the two approaches that rely on SCiFI: either the full SCiFI algorithm (far right), or an approach which fully trains SCiFI but then discards the neural network structure-function map for a linear model.
    Light blue denotes the Metropolis-Hastings algorithm which is function-informed but linear. 
    (a) For butyrate prediction in the gut community experiments of Ref.~\cite{Clark2021}, SCiFI outperforms all other models except one which uses PCA and a neural network. This indicates nonlinearity is important, and that the dominant modes of covariance in abundance data are reflective of butyrate producers.
    (b) Group overlap (confusion matrices) between SCiFI and the Metropolis-Hastings algorithm (left) and between SCiFI and the co-occurrence network (right). Entry $i,j$ in the matrix shows the number of species that were assigned to group $i$ using one method and $j$ using the other method. In this dataset, Metropolis Hastings recovers almost the same groups as SCiFI despite not making accurate predictions.
    (c-d) Same as (a-b) but for succinate prediction.
    (e-f) Same for the Tara Oceans dataset predicting nitrate.
    (g-h) Same for the prediction of nitrate reduction dynamics in soils.
    SCiFI is the only method that either matches or outperforms all other methods across every dataset.
    }
    \label{si_fig:fig3_method_comparison}
\end{figure}

\begin{figure}
    \centering
    \includegraphics[width=\linewidth]{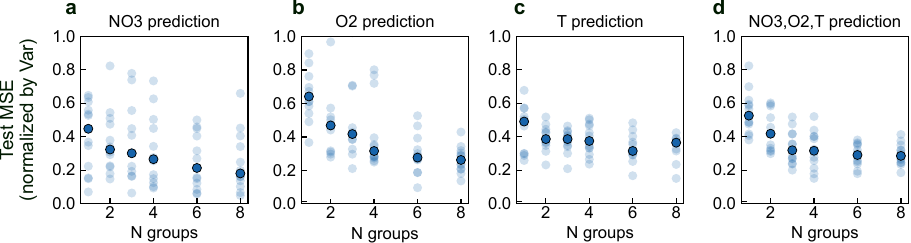}
    \caption{\textbf{Error dependence on number of groups for Tara Oceans metagenome}
    Each panel shows the evolution of mean squared error (MSE) on a held-out test set for varying numbers of groups.
    MSE is normalized by the variance of the predicted function, so that a value of 1 corresponds to the error of a model which simply predicts the mean function value.
    As in Fig.~\ref{fig:validation}, each faint dot corresponds to one model trained on one subset of the training data. Solid dots are the median over the ensemble of models.
    (a) Results when training SCiFI to predict nitrate concentration. Although the median decreases steadily with increasing numbers of groups, the variance is very large.
    (b) Results when predicting oxygen concentration. As with nitrate, the variance of model performance is large.
    (c) Results when predicting temperature.
    (d) Results when simultaneously predicting nitrate, oxygen, and temperature. 
    }
    \label{si_fig:tara_err_vs_nclust}
\end{figure}

\begin{figure}
    \centering
    \includegraphics[width=1.0\linewidth]{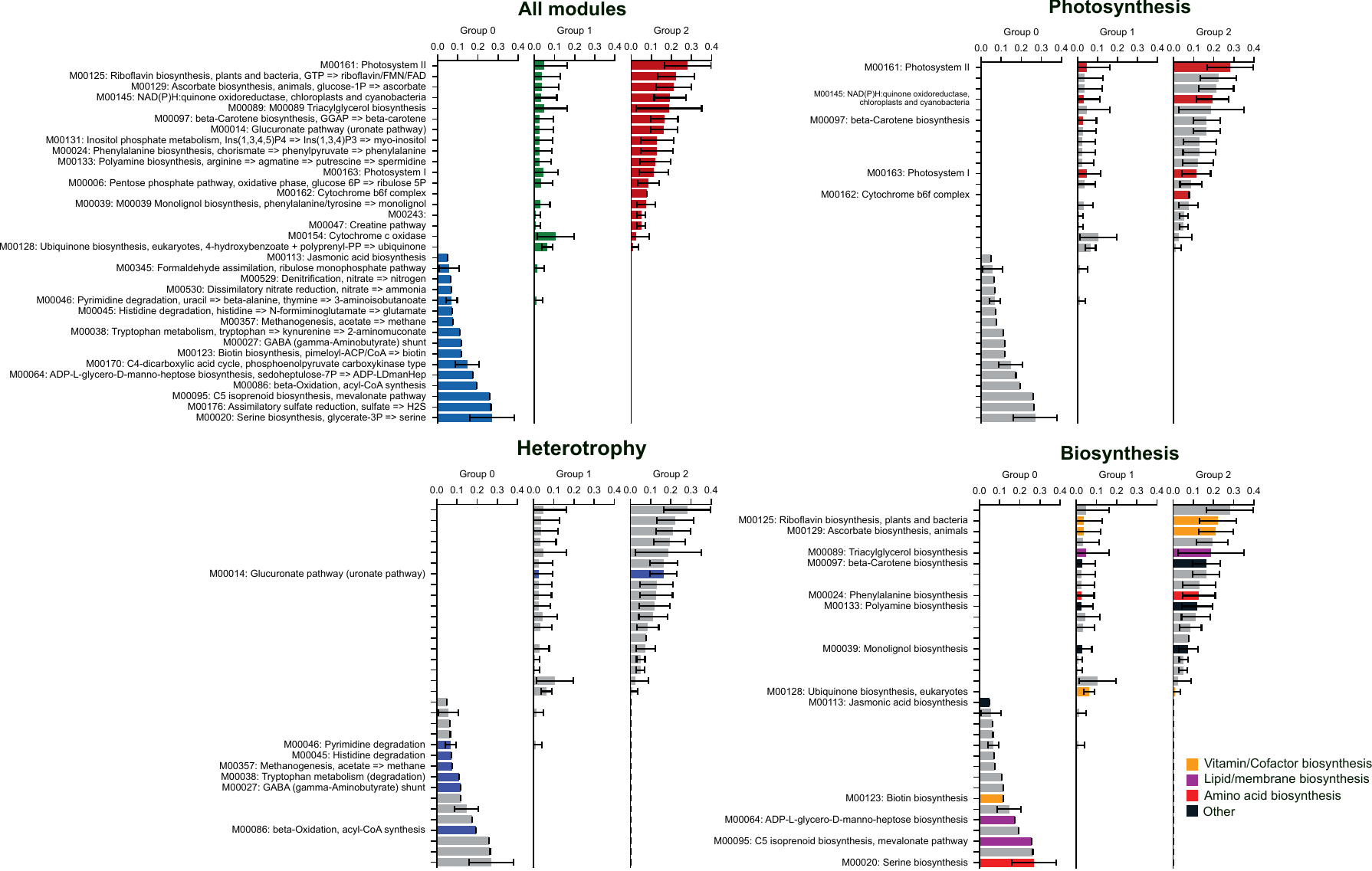}
    \caption{\textbf{Module-level cluster assignments for Tara metagenomics.}
    Here we highlight individual modules isolated using function-driven clustering. 
    Top left shows all KEGG modules with their full descriptions.
    The bar length is the average abundance of the module assigned to each cluster. This average is taken over samples, and over an ensemble of models (we compute the average only over the top 50\% of models). Error bars show the standard deviation (truncated at 0) across the ensemble of models.
    Top right shows how modules related to photosynthesis are concentrated in group 2. 
    Bottom left shows modules related to heterotrophy, which are mostly concentrated in group 0 with the exception of M00014 (Glucuronate pathway). Glucuronate is a precursor for building of extracellular polymeric substances (in bacteria) \cite{decho1990microbial,decho_microbial_2017}. It may also serve as a precursor to ascorbate in animals, suggesting that this may signal the presence of genetic material from eukaryotic plankton which use the protective effects of ascorbate \cite{wheeler_evolution_2015}.
    Bottom right shows modules related to biosynthesis. These are distributed across groups 1 and 3. The surface group (group 3) contains modules used for protective structures (ascorbate, beta carotene, polyamines) and energetically-expensive structures (phenylalanine, monolignol).
    Riboflavin biosynthesis is important in the surface group because riboflavin degrades rapidly due to light \cite{zika_photochemistry_1987}, meaning it must be constantly generated to maintain the constant supply necessary for growth.
    The deep group (group 1) contains structures used in archea (Mevalonate, ADP-heptose), simple amino acids (serine) and cofactors which are potentially useful for hetertrophy (biotin, used for anaplerosis).
    }
    \label{si_fig:tara_modules}
\end{figure}

\begin{figure}
    \centering
    \includegraphics[width=0.9\linewidth]{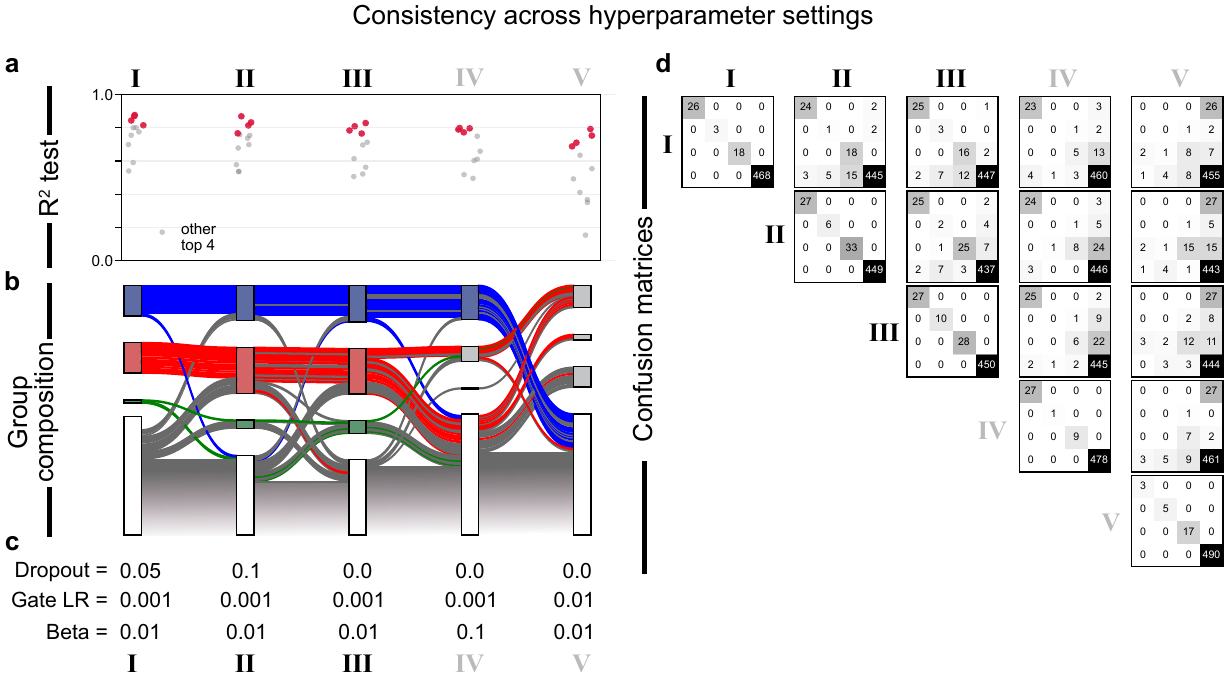}
    \caption{\textbf{Consistency of groups across hyperparameters: Tara metagenomics.}
    Here we evaluate the sensitivity of the learned gene groupings from Fig.~\ref{fig:tara_metagenomics} to training hyperparameters.
    (a) $R^2$ values for each of the different hyperparameter sets, enumerated with Roman numerals. Small dots correspond to single models in the ensemble. The top four models highlighted in red are used to determine the consensus grouping (Methods).
    Performance is mostly constant for hyperparameter sets I-III but begins to degrade in IV and V.
    (b) Consensus group composition resulting for each hyperparameter set. 
    Each set of nodes (rectangles) in the flow diagram are aligned vertically with the corresponding hyperparameter set from (a). Each bar between nodes represents one gene module; its thickness corresponds to its average abundance.
    The genes in each of the three groups from Fig.~\ref{fig:tara_metagenomics} are colored accordingly.
    The bottom group (gray) represents all genes that are removed by the gating procedure. These are not used as an input to the structure-function map neural network.
    (c) The hyperparameters corresponding to each set I-V.
    (d) Confusion matrices to compare the group assignments across hyperparameter sets. Each matrix is a comparison between one set of hyperparameters (Roman numerals).
    Entry $i,j$ in the matrix is the number of genes that are assigned to group $i$ for one hyperparameter set and to group $j$ for another hyperparameter set.
    }
    \label{si_fig:fig4_metag_hparam_consistency}
\end{figure}

\begin{figure}[tp]
    \centering
    \includegraphics[width=0.67\linewidth]{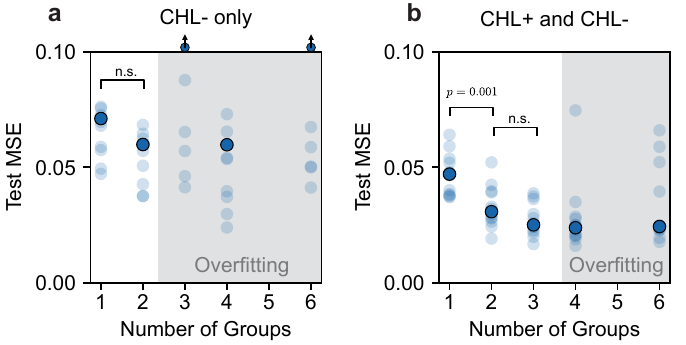}
    \caption{\textbf{Test (mean squared) error versus number of groups for soil data.}
    Predictions are made with ASV-level abundance data as input, using only CHL-untreated samples (a) or both treated and untreated samples (b). Models are trained with the gating procedure described in Methods. Each semi-transparent point is one model trained with a different test-train split of the data. Solid point is the median loss. 
    There are clear signs of overfitting for larger numbers of groups (gray background) because test error begins to grow rapidly. 
    Note that some models with test MSE > 0.10 are not shown. Cases where the median MSE is greater than this value are shown with a point lying on top of the axis in (a). 
    Parameters $lr=10^{-3}$, $lr_{\text{gate}}=10^{-1}$, $\beta=10^{-2}$, $p_\text{dropout}=0.05$. $p$-values shown for two-sided t-test comparing the means of the distribution of errors. Other differences are not significant (n.s.); for CHL- only the differences between 1 and 2 groups has $p=0.28$ and for both treated and non-treated the difference between 2 and 3 groups is $p=0.10$. 
    }
    \label{si_fig:soil_err_vs_nclust}
\end{figure}

While SCiFI is continuously adjusting $P$ behind the scenes, we still ensure that the neural network only ever sees ``hard'' clusterings during training. 
To do this we sample a hard clustering $C$ using the above matrix $P$: for every species (row), a group assignment (one hot vector) is generated by sampled from the distribution in the corresponding row of $P$ (bottom of Fig.~\ref{si_fig:gumbel_demo}). 
The Gumbel-softmax trick achieves this by randomly drawing a matrix $\Gamma\in\mathbb{R}^{n\times N_{\text{clust}}}$, where each entry is Gumbel distributed, and with it computing 
\begin{align}
    C=\text{softmax}\left(\frac{1}{\tau}(\Gamma + \log P )\right)
\end{align}
where the softmax is taken over rows (i.e. each row is normalized to sum to 1) \cite{gumbel_maddison_2014, gumbel_maddison_2017,gumbel_jang_2017}. 
The matrix $C$ is a sample of a hard clustering, consisting of approximately zeros and ones (the approximation is better as $\tau\rightarrow 0$).
The relationship between $C$ and $P$ is shown for two different values of $\tau$ in Figs.~\ref{si_fig:gumbel_demo} and \ref{si_fig:gumbel_demo_in_time}. In particular, Figure~\ref{si_fig:gumbel_demo_in_time} shows how, while $P$ evolves slowly in time to the correct clustering, $C$ undergoes large changes due to the stochastic sampling which enables SCiFI to explore many possible clusterings.

\subsubsection{Step-by-step guide to training}
Here we present a step-by-step guide that explains the full training procedure in detail.

\noindent \textbf{Step 1: Data preparation.} The data should consist of pairs $\{x_i, f_i\}$ where $x_i\in\mathbb{R}^{n}$ is a single (row) vector of abundances and $f_i\in\mathbb{R}^{m}$ is the corresponding vector of functional measurements. By assembling individual samples into a matrix we have $X\in\mathbb{R}^{N\times n}$ $F\in\mathbb{R}^{N\times m}$ where $N$ is the number of samples in the dataset.
In all of the examples in this work we perform cross validation by choosing $N_{\text{ens}}$ (ens $=$ ``ensemble'') different test/train splits of the dataset. We typically choose $N_{\text{train}}=0.8\cdot N$ training samples with the remaining 20\% used for testing. The 80\% of samples used are randomly drawn from the dataset. We use an ensemble size of $N_{\text{ens}}=12$ to 15. Let us write the set of training data for each ensemble member as $\{X^{\text{train}}_{\mu}, F^\text{train}_{\mu}\}$ where $\mu=1..N_{\text{ens}}$, and analogously for the test data.

\noindent \textbf{Step 2: SCiFI Training.} 
We train one model per test/train split $\{X^{\text{train}}_{\mu}, F^\text{train}_{\mu}\}$, resulting in an ensemble of $N_{\text{ens}}$ models.
The neural network used here are small to mitigate overfitting, typically 3 layers with 128 neurons each. The precise width and depth do not strongly impact model performance.
The key hyperparameters which impact performance are the learning rate $\eta$ and the ``temperature'' $\tau$. We typically allow $\tau$ to reduce during training via an annealing schedule which decays exponentially from $\tau_{\text{max}}$ (typically 1.0) to $\tau_{\text{min}}$ (typically 0.1) at a rate $r_{\tau}$. To determine the rate $r_{\tau}$ and the learning rate $\eta$ we do a hyperparameter gridsearch over a handful of possible values for each. When gating is used, that introduces another hyperparameter $\beta$ which is also included in this gridsearch. For a given set of hyperparameters we train the entire ensemble of models with their respective training data. Training is interrupted either due to the test loss ceasing to decrease (early stopping) or when a maximal number of iterations is reached.

\noindent \textbf{Step 3: Post-processing and group extraction.} 
The training procedure above generates an ensemble of $N_{\text{ens}}$ clusterings. From this point on, we work with the soft clusterings $\{P_\mu\}_{\mu=1..N_{\text{ens}}}$.
Due to stochasticity which arises both from the different training data splits and the Gumbel sampling trick, the matrices $P_\mu$ may show differences. 

\begin{figure}[tp]
    \centering
    \includegraphics[width=1.0\linewidth]{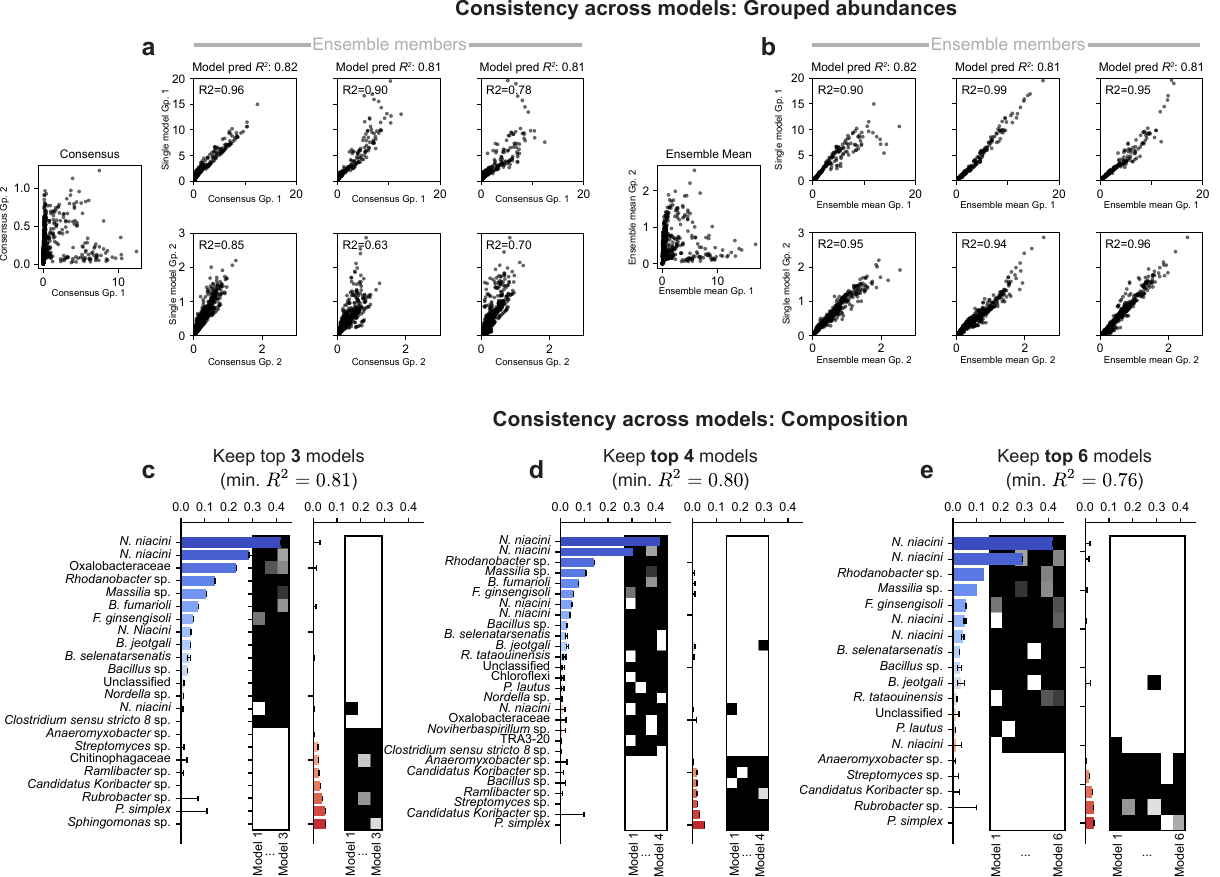}
    \caption{\textbf{Learned assignments are consistent across different selection criteria.}
    We train an ensemble of models, each of which may learn a different grouping. 
    We explore two ways of determining ``average'' group abundances. 
    First, we can take a ``consensus grouping'' which selects all species that are placed into a single group more than 50\% of the time across the ensemble (see Methods). 
    Second, we can first compute the group abundances for each ensemble member individually, and then compute the average across the ensemble.
    These give different results because in the first approach one neglects small or unimportant groups, while in the second they are taken into account because they contribute to the group abundance for each individual model.
    (a) Comparison of group 1 and 2 abundances (top and bottom row) for the individual models compared to the consensus group.
    (b) Comparison of individual model group abundances to the ensemble mean abundances.
    (c-d) Show how our selection of the consensus group members is stable whether we take the top 25\% (3) of models or the top 50\% (6) to build the consensus group. From this subset of models, a group member is included if it is present in 75\% or more of the individual models. 
    (c) Left axis shows the average end-point abundances (blue) of each species included in the consensus group. The inset matrix shows the presence of each species across all models.
    Right axis shows abundances (red) of species in Group 2. 
    These species are all those included if we determine our consensus group from the top 3 performing models.
    (d) Consensus group members derived from the top 4 performing models, sorted by abundance (blue, red) with their presence/absence across models (inset matrices).
    (e) Consensus group members derived from the top 6 performing models. 
    }
    \label{si_fig:soil_group_consistency}
\end{figure}

We extract a ``consensus'' clustering as follows. First, we sort the ensemble of models based on their predictive accuracy ($R^2$ of their function predictions versus the true predictions).
Next, we choose a subset of high-performing models, typically the top 25\% to 50\%.
In SI Fig.~\ref{si_fig:soil_group_consistency} we show how in the soil data, the consensus grouping is insensitive to the choice of how many top-performing models we take.
We then make the consensus grouping based on majority rule: i.e. if five ensemble members assign species $i$ to group 1 and four assign it to group 2, then the consensus assignment will be to group 1. 
When gating is used, the above procedure is modified slightly by adding the additional condition that species are only included in the consensus grouping if they are not gated in a majority of ensemble members.
Clusterings are only unique up to a permutation invariance which we must account for. In other words, the groups $\{1,2,3\}$ for one model may correspond to the groups $\{2,3,1\}$ in another model just because of the arbitrariness of the group labeling. 
We correct for this by reordering groups based on their average abundance, or, in the case of the ocean metagenome, based on their correlation coefficient with nitrate.

\subsection*{Simulated data}
To test SCiFI we construct a simulated dataset where community function arises as an effectively nonlinear combination of group abundances. 
Such an effective nonlinearity may come about even from a structure-function mapping with a linear form if the linear coefficients vary according to some hidden parameter.
This is inspired by datasets which aggregate samples from environments which vary either naturally or due to experimentally-applied perturbations. 
For example, this may occur in communities which modulate their behavior depending on pH, which is directly relevant for the gut and soil datasets we consider later.

We consider a simulated dataset consisting of 3 groups assembled from 12 species.
The function depends linearly on the group abundances, 
\begin{align}
    f(x)=\sum_{i=1}^{3} a_iG_i(x)=\sum_{i=1}^{3} a_i\sum_{j\in I_i}x_j
    \label{methods_eq:simulated_data_f}
\end{align}
where $\vec{a}$ is a vector of parameters and $I_i$ is the set of species indices that are in group $i$.
To model the effect of varying environment (either due to perturbation or naturally-occuring), we let the parameter vector $a$ vary across three ``regimes'':
\begin{align}
    \vec{a}^{(1)}&=(0.1,1,1),\quad\text{when }(G_1>\text{med}(G_1))\,\land\, (G_2>\text{med}(G_2))
    \\
    \vec{a}^{(2)}&=(1,-0.5,-0.1),\quad\text{when }(G_1>\text{med}(G_1))\,\land\, (G_2<\text{med}(G_2))
    \\
    \vec{a}^{(1)}&=(-1,0.1,-1),\quad\text{when }(G_1<\text{med}(G_1))
\end{align}
where $\text{med}(G)$ denotes the median abundance of group $G$.
Although the functional form of $f$ is linear in Eq.~\eqref{methods_eq:simulated_data_f}, globally it is nonlinear due to the dependence of the parameters $\vec{a}$ on regime. 
\subsubsection*{Evaluation with Jaccard Index}
Because the true groupings are known in this synthetic example, we can evaluate the quality of the group recovery using the Jaccard index. 
The Jaccard index compares the overlap (intersection) between two groupings with their union. To compare group $i$ from one grouping with group $j$ from another, one computes
\begin{align*}
    J_{ij}=\frac{\text{$N_{\text{species}}$ in group $i$ AND group $j$}}{\text{($N_{\text{species}}$ in group $i$) + ($N_{\text{species}}$ in group $j$)}}
\end{align*}
If rows of $J$ correspond to the ground truth groups, then the group recovery for group $i$ is given by $J_i = \max_jJ_{ij}$. The total Jaccard index is given as the average across all groups, $J=\frac{1}{G}\sum_{i=1}^GJ_i$ where $G$ denotes the number of groups.

\subsection{Mathematical Modeling}
For the mathematical modeling in Fig.~\ref{fig:soil_CR_dynamics} we start from the approach used in Ref.~\cite{lee2024functional}. 
Our model (Eq.~\ref{eq:consumer_resource}) is a minimal two-biomass extension of the single-biomass consumer resource model introduced in that work. The full dynamical system is given by
\begin{align}
    \dot x_1 &= \gamma_1r_1x_1\frac{A}{A+K_A}\frac{C}{C+K_C}
    \label{si_eq:x1}
    \\
    \dot x_2 &= \gamma_2r_2x_2\frac{A}{A+K_A}\frac{C}{C+K_C}
    \label{si_eq:x2}
    \\
    \dot A &= -(r_1x_1 + r_2x_2)\frac{A}{A+K_A}
    \label{si_eq:A}
    \\
    \dot C &= -(x_1 + r_Cx_2)\frac{C}{C+K_C}.
    \label{si_eq:C}
\end{align}
The carbon consumption equation starts from a similar form to that for nitrate $A$, but we absorbed the rate that multiples $x_1$ by rescaling $C$ so that there is only one remaining parameter $r_C$.
The full set of parameters to be determined are $\{\gamma_1, \gamma_2, r_1, r_2, r_C, C_0, K_A, K_C\}$, where $C_0$ is the initial carbon abundance; we treat the $\{\gamma_1, \gamma_2, K_A, K_C, r_C\}$ as parameters which are fixed across all conditions, while the rates $\{r_1,r_2,C_0\}$ may vary from sample to sample which reflects how reduction rates and carbon availability change with pH \cite{lee2024functional}. 
The goal is to optimize the parameters so that the above ordinary differential equation (ODE) correctly predicts the end-point biomasses and the evolution of nitrate concentration
$\{x_1(T), x_2(T), A(t_1),...,A(T)\}$ starting from the (experimentally measured) initial conditions $\{x_1(0), x_2(0), A(0)\}$. 

This is an instance of an ODE-constrained optimization problem. While the parameters may be found with e.g. the adjoint method, this approach is computationally slow. We instead rely on the fact that the dynamics is approximately given by an analytic solution. As long as the nutrients are not depleted ($A>0$ and $C>0$) the growth is approximately exponential, as is the depletion of $A$ and $C$. When one of these is depleted, growth ceases so that $x_1=$const, $x_2$=const, and $A$ decays linearly in time. 
In this procedure, we can use the analytical solution and search for the depletion time $t^*$ instead of the initial carbon $C_0$. 
The results are insensitive to the affinities $K_A$ and $K_C$ and so we fix them to some small value ($10^{-3}$). We find the remaining global parameters using a grid search over a range of possible hyperparameters, and find $\gamma_1=0.4$, $\gamma_2=2.44$ and results which are insensitive to $r_C=1$. More information about this approach may be found in Ref.~\cite{lee2024functional}.

\subsubsection*{Null model comparison}
\begin{figure}[tp]
    \centering
    \includegraphics[width=0.85\linewidth]{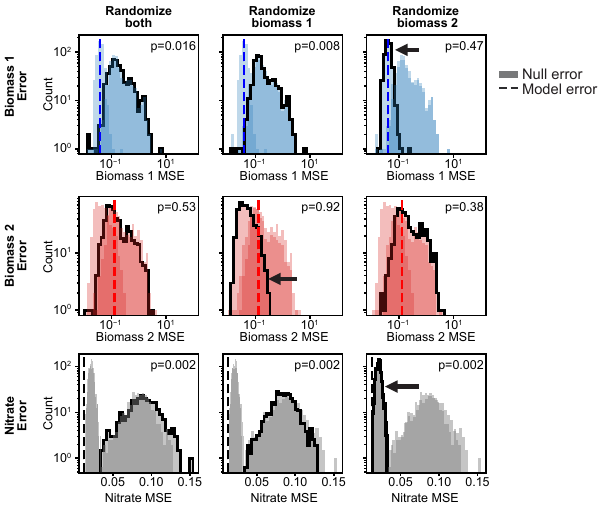}
    \caption{\textbf{
    Null model comparison.}
    A full description of the figure is found in Methods.
    Each row shows the distribution (highlighted in black) of errors using a null model of group abundances which draws either both groups from a null distribution (left column), only group 1 (middle) or only group 2 (right). Distributions from other null models are shown, not highlighted, for comparison.
    Top row shows error in biomass 1 end-point abundance prediction; middle shows error for biomass 2; bottom shows error for nitrate dynamics. $p$-values shown are the empirical probabilities of observing a null model error that is lower than the correct model error (shown with vertical dashed line).
    The arrows highlight some observations which are discussed in Methods.
    }
    \label{si_fig:soil_null_models}
\end{figure}

\begin{figure}[tp]
    \centering
    \includegraphics[width=0.9\linewidth]{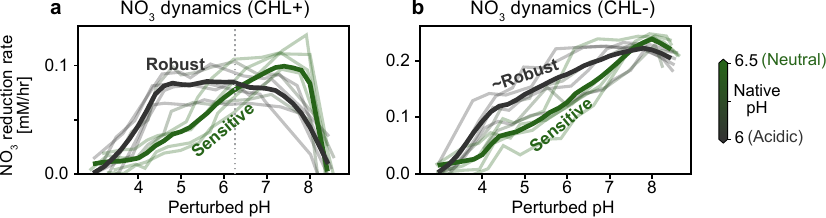}
    \caption{\textbf{Nitrate reduction dynamics in CHL treated and untreated conditions.}
    (a) Reproduction of Fig.~\ref{fig:soil_genomics}c for comparison. 
    (b) Nitrate reduction rate for each soil sample as a function of perturbed pH. As in Fig.~\ref{fig:soil_genomics}c, each soil sample (faint lines) is colored by native pH. Solid lines show the average across acidic (pH < 6.25) and neutral (pH > 6.25) soils.
    The difference between robust and sensitive soils is less pronounced than in the CHL-treated samples but still present.
    }
    \label{si_fig:no3_reduction_chlplus}
\end{figure}

To evaluate whether our groups are the ``correct'' variables in the mathematical model, we compare the performance of the fit model to the performance of one fit using randomized groupings.
Random groups are generated as follows: For a particular group, say Group $i$, we construct an assembly of species whose total initial abundance is approximately equal to the initial abundance of Group $i$. We can randomize just a single group, or both groups, after which we apply the same fitting procedure as used for the consumer resource model in the main text.
The result of this procedure is shown in SI Fig.~\ref{si_fig:soil_null_models}.

We consider the randomization of both groups, only group 1, or only group 2; each case corresponds to one column in Fig.~\ref{si_fig:soil_null_models}.
Each panel shows the distribution of errors of the null model. Top row shows biomass 1 (mean squared) error, middle row shows biomass 2 error, and bottom row shows the error in nitrate prediction. 
Each plot highlights with a black outline the distribution that arises from shuffling the variable denoted in that column; the  distributions from the other randomization procedures are not highlighted but shown in all panels for comparison.
We compare the highlighted distributions to the model error using the ``correct'' groups (shown as vertical dashed line) by computing the $p$-value shown in the upper right corner.

We find that shuffling either of the biomasses individually or both together leads to significantly worse performance than the true model (bottom row). However, shuffling biomass 2 only, while still worse than the true model, degrades performance to a lesser extent than the other two randomization procedures (horizontal arrow in bottom row). There is also almost no effect on the prediction of biomass 1 (arrow in top row).
Shuffling biomass 1 significantly degrades predictive power for both nitrate dynamics and biomass 1 end-point abundance. However, the performance in biomass 2 prediction actually increases (arrow in middle row), possibly hinting that it could not be as accurately predicted because of a trade-off in the model's ability to predict both biomass 1 and biomass 2.
Together, these results suggest that biomass 2 is not as important as biomass 1 in making predictions of nitrate dynamics or biomass 1 dynamics.

\subsection*{Microbiome datasets}
\subsubsection*{Human gut synthetic community dataset}
Synthetic community experiments with human gut microbial strains were done by Clark \textit{et al.} (2021), where they inoculated 1850 different combinations of 26 representative gut strains in the media in the anaerobic chamber to measure the final concentrations of produced butyrate, succinate, acetate, and lactate. For each synthetic community, the relative abundance of the endpoint community was measured by 16S rRNA amplicon sequencing as described by the original authors~\cite{Clark2021}. We used the endpoint relative abundance of strains as compositional input and produced butyrate and succinate concentrations as functional output for training the neural network. The gene annotation information of each strain's genome (Biocyc database) from the original authors~\cite{Qian2025} was used to validate the functional clusters by enzyme presence or absence. The dimension (sample × strain) of the compositional data was 1850$\times$26. The functional data used for neural network prediction had a dimension (sample$\times$metabolite concentration) of 1850$\times$1, respectively for butyrate and succinate.

\subsubsection*{Tara Oceans natural community dataset}
136 Tara Oceans samples were sampled by the original authors~\cite{Louca2016, Sunagawa2015} across the oceans at different depths. For each sample, they measured nitrate concentration, temperature, and dissolved oxygen level (6 samples have NaN values for at least one of these measurements and are omitted), which we used as functional outputs to train the neural network. 
The relative abundance of taxa in each sample was derived from 16S rDNA fragments in the metagenomic sequencing (miTAGs) as described by the original authors~\cite{Louca2016, Sunagawa2015} and Shan et al~\cite{Shan2023}. The dimension (sample$\times$taxa) of the compositional data was 136$\times$97 at the genus level.

For the gene module abundance of each sample, we used the annotated gene tables uploaded by the original authors~\cite{Louca2016, Sunagawa2015}. The dimension (sample$\times$gene) of the gene composition data was 136$\times$515 for module-level annotation. The functional data used for neural network prediction had a dimension (sample × environmental variable) of 136$\times$1 for the case we identify functional groupings with nitrate concentrations or 130$\times$3 for the case we use all 3 environmental variables (nitrate concentration, temperature, and dissolved oxygen level) to identify functional groups, as we removed 6 samples without corresponding oxygen measurements.

\subsubsection*{Soil pH perturbation dataset}
20 Topsoils were collected across a native pH gradient (4.7--8.32) in the Cook Agronomy Farm (CAF) in the Long-Term Agroecosystem Research (LTAR) (Pullman, WA, USA) and perturbed each soil slurry into 13 different pH levels as well as adding 2mM nitrate to initiate denitrification in 1:2 soil-to-water slurries under anaerobic conditions, as described previously.~\cite{lee2024functional}. To infer indigenous biomass activity of nitrate reduction, we included chloramphenicol-treated (CHL+) controls, in which protein synthesis and growth were inhibited, yielding linear nitrate reduction dynamics. In contrast, untreated controls (CHL–) allowed microbial growth and physiological responses to the imposed pH conditions. For functional outputs, we measured the functional dynamics (10 time points during the 4-day incubation) of nitrate, nitrite, and ammonium concentrations. The absolute abundance of each taxon per sample was acquired by dividing reads from 16S rRNA amplicon sequencing with reads from the internal standard of known spiked-in strains~\cite{lee2024functional}. 16S rRNA amplicon sequencing was performed on endpoint perturbed samples derived from 10 of the 20 CAF sampled soils.

We removed 18 samples lacking complete triplicate CHL+/CHL- pairs in Soil12 (perturbed unit 20 and 40) and Soil15 (perturbed unit 12). We removed taxa that had fewer than 1,000 total reads across the entire sequencing dataset to minimize noise and sparsity coming from very low-abundance taxa. The dimensions (sample × taxa) of the compositional data were 762$\times$28 for phyla, 762$\times$256 for family, and 762$\times$4395 for ASV level. The functional data used for neural network prediction had a dimension (sample × metabolite dynamics) of 762$\times$10 for nitrate, 762$\times$20 for nitrate and nitrite, 762$\times$30 for nitrate, nitrite, and ammonium. 

\subsubsection*{Strain isolation and whole genome sequencing}
For isolation, as described previously~\cite{lee2024functional}, we thawed endpoint slurry samples from soil pH perturbation experiments stored in 25\% glycerol in -80$^\circ$C, streaked onto 1/10× tryptic soy agar (TSA) plates and 1/10x Reasoner’s 2A (R2A) plates adjusted to the endpoint pH of each slurry sample. Plates were incubated at 30$^\circ$C and checked for growth every 24 hours. Genomic DNA was extracted from pure colonies using the DNeasy Ultraclean Microbial Kit (Qiagen, Hilden, Germany). Unique strains were identified by Sanger sequencing of the 16S rRNA region amplified by 27F and 1492R primers, followed by BLAST against the NCBI 16S rRNA reference sequences for taxonomic identification. We decided to further investigate the genetic composition of the two strains taxonomically identified as \textit{Neobacillus fumarioli} (Group 1 representative strain) and \textit{Peribacillus simplex} (Group 2 representative strain) with whole genome sequencing.

For whole-genome sequencing, approximately $5 \times 10^9$ cells were preserved in DNA/RNA Shield (Zymo Research) and submitted to Plasmidsaurus (USA). Genomic DNA was prepared using an amplification-free library construction protocol (ONT V14 chemistry) and sequenced on Oxford Nanopore Technology (ONT) R10.4.1 flow cells. Raw signals were basecalled using the Dorado Super-Accurate (SUP) model with a default Q10 quality filter. Read quality was further refined by removing the lowest 5$\%$ of reads using Filtlong v0.2.1. De novo assembly was executed via the Autocycler pipeline, which generates a consensus from multiple independent assemblers, including Flye v2.9.6+, Hifiasm, and Plassembler v1.8.0+. The final consensus assembly was polished using Medaka v1.8.0 and oriented to the dnaA start position using dnaapler. Structural and functional annotation was performed using Bakta v1.11.4 utilizing the v6.0 full database. Species identification was verified through Mash v2.3 and Sourmash v4.6.1 against RefSeq and GenBank databases, while genome completeness and contamination were assessed via CheckM v1.2.2.

\begin{table}[htp]
    \centering 
    \begin{tabular}{|c|c|c|c|c|c|c|}
    \hline
     Dataset &
     Input & 
     Function & 
     $N_{\text{samples}}$ & 
     $N_{\text{train}}$ & 
     $N_{\text{ens}}$ &
     Figure \\
     \hline\hline
     Gut (synth.) \cite{Clark2021} & 
     Strains ($d=30$) &
     Butyrate ($d=1$) &
     1592 &
     1273 &
     20 &
     \ref{fig:validation}a-d
     \\
     \hline
     Gut (synth.) \cite{Clark2021} & 
     Strains ($d=30$) &
     Succinate ($d=1$) &
     1592 &
     1273 &
     20 &
     \ref{fig:validation}e-h
     \\
     \hline
     Ocean \cite{Sunagawa2015} & 
     Genera ($d=97$) & NO$_3$ ($d=1$) &
     136 & 108 & 15 & \ref{fig:validation}i-l \\
     \hline
     Soil \cite{lee2024functional} & 
     Phyla ($d=28$) &
     NO$_3$(t) ($d=10$) &
     381 &
     304 &
     15 &
     \ref{fig:validation}m-p
     \\
     \hline
     Ocean \cite{Sunagawa2015} & 
     Gene modules ($d=515$) & (NO$_3$,O$_2$,T) ($d=3$) &
     130 & 104 & 12 & \ref{fig:tara_metagenomics} \\
     \hline
     Soil \cite{lee2024functional} & 
     ASVs ($d=4395$) & NO$_3$(t) ($d=10$) &
     762 & 570 & 12 & \ref{fig:soil_CR_dynamics},\ref{fig:soil_genomics} \\
     \hline
    \end{tabular}
    \caption{\textbf{Dataset overview}
    Summary of each dataset used in this work. Input states how abundance data is defined, with $d$ the dimension (number of species/genes). 
    Function states the functional data and its dimensionality $d$.
    The total number of samples in the dataset is given by $N_{\text{samples}}$, and the number of samples used for training is $N_{\text{train}}$.
    The number of samples in the two rows of the ocean dataset differ because some values of oxygen concentration are missing, which means we omit these samples.
    The number of samples differs between the two soil datasets because in the first case only non-treated samples are used, while in the second both chloramphenicol-treated and non-treated samples are used.
    The number of models trained in the whole ensemble is given by $N_{\text{ens}}$; for details see the ``Step-by-step guide to training'' in Methods. 
    }
    \label{tab:data_summary}
\end{table}

\subsection*{Data availability}
All data used to train the neural networks will be made available upon publication.

\subsection*{Code availability}
Code deploying our method on a synthetic dataset may be found in \url{github.com/schmittms/function_informed_clustering}. 
All code used to generate the figures will be made public upon acceptance for publication.

\section*{Acknowledgements}
The soil-related research was conducted as part of the Long-Term Agroecosystem Research (LTAR) Network, which is supported by the United States Department of Agriculture. We thank Otto Cordero for constructive discussions and feedback. We acknowledge Jocelyn Wang for soil strain isolation.

This research was supported in part by grants from the NSF (DMS-2235451) and Simons Foundation (MPS-NITMB-00005320) to the NSF-Simons National Institute for Theory and Mathematics in Biology (NITMB).
S.K. acknowledges the Center for the Physics of Evolving Systems at the University of Chicago, National Institute of General Medical Sciences R01GM151538.
S.K. acknowledges a CAREER award from the National Science Foundation (BIO/MCB 2340416). 
V.V. and S.K. acknowledge support from the National Science Foundation through the Physics Frontier Center for Living Systems (PHY2317138). 
V. V. is a Chan Zuckerberg Biohub Chicago Investigators.
Any opinions, findings, conclusions, or recommendations expressed in this material are those of the authors and do not necessarily reflect the views of the National Science Foundation. S.K. acknowledges support from the Army Research Office (AWD106455).
This work was completed in part with resources provided by the University of
Chicago’s Research Computing Center.

\bibliographystyle{unsrt.bst} 
\textnormal{\bibliography{bib.bib}}

\clearpage

\vspace*{\fill}
\section{Supplementary Information}
\vspace*{\fill}

\clearpage

\clearpage

\clearpage

\clearpage

\clearpage

\clearpage

\clearpage

\end{document}